\documentclass[sigconf]{acmart}
\usepackage{booktabs} 
\usepackage{graphicx}
\usepackage{fancybox}
\usepackage{balance}

\newcommand{\hypobox}[1]{

        \begin{center}\noindent\thicklines\setlength{\fboxsep}{8pt}\cornersize{0.2}\ovalbox{

                \begin{minipage}{3.0in}

                        \textit{#1}

                \end{minipage}} 

        \end{center}} 
        
\setcopyright{rightsretained}

\begin{document}
\title{Which Pull Requests Get Accepted and Why? A study of popular NPM Packages}

\author{Tapajit Dey}
\email{tdey2@vols.utk.edu}
\affiliation{%
  \department{Electrical Engineering and Computer Science}
  \institution{The University of Tennessee}
  \streetaddress{1520 Middle Dr.}
  \city{Knoxville}
  \state{TN}
  \country{USA}
  \postcode{37996}
}
\author{Audris Mockus}
\email{audris@mockus.org}
\affiliation{%
  \department{Electrical Engineering and Computer Science}
  \institution{The University of Tennessee}
  \streetaddress{1520 Middle Dr.}
  \city{Knoxville}
  \state{TN}
  \country{USA}
  \postcode{37996}
}


\begin{abstract}
Background: Pull Request (PR) Integrators often face challenges in
terms of multiple concurrent PRs, so the ability to gauge which of
the PRs will get accepted can help them balance their workload. PR
creators would benefit from knowing if certain characteristics of their PRs
may increase the chances of acceptance.
Aim: We modeled the probability that a PR will be accepted within a
month after creation using a Random Forest model utilizing 50
predictors representing properties of the author, PR, and the
project to which PR is submitted. 
Method: 483,988 PRs from 4218 popular NPM packages were analysed and
we selected a subset of 14 predictors sufficient for a tuned Random
Forest model to reach high accuracy. 
Result: An AUC-ROC value of 0.95 was achieved predicting PR
acceptance. The model excluding PR properties that change after
submission gave an AUC-ROC value of 0.89. We tested the utility of
our model in practical scenarios by training it with historical data
for the NPM package \textit{bootstrap} and predicting if the PRs
submitted in future will be accepted. This gave us an AUC-ROC value
of 0.94 with all 14 predictors, and 0.77 excluding PR properties that
change after its creation.
Conclusion: PR integrators can use our model for a highly accurate
assessment of the quality of the open PRs and PR creators may
benefit from the model by understanding which characteristics of
their PRs may be undesirable from the integrators' perspective. The
model can be implemented as a tool, which we plan to do as a future work.
\end{abstract}

%
%



\maketitle

\section{Introduction}\label{s:intro}
The source code of an Open Source Software is publicly available and
can be modified and reused with limited restrictions by the
public. A major upside of this is that the users
can and do contribute to and reuse the software. User contributions
consist of primarily reporting and fixing bugs and adding or
requesting for additional features to the software. The Pull Request
based development model, that is prevalent among most OSS version
control systems, is a distributed development model that allows
other user-developers to make contributions, which can be easily tracked,
thus supporting the collaboration between the maintainer and the contributor. The
mechanics of the approach is that potential contributors first fork
(clone) the original repository they are planning to contribute to,
and after making the code changes they create a Pull Request (PR),
which is essentially an issue with a patch included. Then a
maintainer of the original project (the integrator) inspects the
PR's code changes and decides if it can be merged to the project and
interacts via discussion thread associated with the PR, with the
submitter and other maintainers and/or potential contributors. 

This development model allows developers outside of a project to
contribute while not compromising the code of the original project
by only merging approved changes to the repository and was found to
be associated with shorter review times and larger numbers of
contributors compared to mailing list code contribution
models~\cite{zhu2016effectiveness}. However, inspecting PRs is a
crucial task requiring a significant amount of effort from the
integrator~\cite{rigby2014peer}. It has been extensively documented (see,
e.g.~\cite{xie2013impact}) that large numbers of low-quality issues
may overwhelm the projects and the same is true for PRs. It was
reported in~\cite{gousios2015work}, based on a survey of 750
integrators from high volume projects, that the top two challenges
they face when working with pull requests are\textit{ maintaining
  project quality} and \textit{prioritizing work in the face of
  multiple concurrent pull requests}. Therefore, having a good
estimate of PR quality would be highly desirable and would help the
integrator to prioritize the PRs.  Moreover, the creators of the
pull requests would also benefit from having an indication of the
quality of a PR they are about to submit, and knowing what factors
affect the quality of the PR could help them improve that PR and the ones
they might submit in future. E.g. if they find that having too many
commits in their PR is affecting its quality, they might try
submitting smaller patches with fewer commits.

However, PR ``quality'' has no universal definition, it is a highly
contextual factor and might mean very different things in different
scenarios. To operationalize the measure of quality for PRs, we chose
the ultimately pragmatic indicator: whether it is merged (accepted)
or not. After all, it should be based on the contextual knowledge of
the integrator at the time of acceptance and should take into
account a variate of factors the integrator has to consider when
accepting a PR. A comprehensive treatment of code contribution
theory in Rigby et. al.~\cite{rigby2014peer} considers the acceptance
rate as one of the most fundamental properties of the peer-review
systems. In this paper we, therefore, define the quality of a PR by
its probability of getting merged.

To conduct an empirical study investigating PR quality we chose node
package manager (NPM) because of the size of the associated ecosystem and availability of data. NPM is a package manager of JavaScript packages, and is one of the largest OSS communities at present, with over 932,000 different packages (Apr, 2019) and millions of users (estimated 4 million in 2016~\cite{npmuser}, and about 4000 new users on an average day\footnote{https://twitter.com/seldo/status/880271676675547136}). NPM is used heavily by companies as well. According to the NPM website\footnote{https://www.npmjs.com/}, all 500 of the Fortune 500 companies use NPM, and they claim that: `` \textit{Every company with a website uses npm, from small development shops to the largest enterprises in the world.}'' However, most packages in NPM are not widely used and have limited or no issues or PRs. We, therefore, focused on 4218 NPM packages with over 10,000 monthly downloads (the ``popular'' packages) since January, 2018, that also has an active GitHub repository with at least one Pull Request. All 483,988 PRs, that were filed against these packages (until January, 2019 when the data was collected), were obtained using the GitHub API. 

A previous study~\cite{prioritizer} described a priority inbox type of approach for Pull Request prioritization aimed at helping integrators deal with multiple concurrent PRs. They introduced a prototype tool called PRioritizer, a service-oriented architecture built on top of GHTorrent~\cite{Ghtorrent}, that examines all open pull requests and presents the project integrators with the top pull requests that potentially need their immediate attention.  They split the time into configurable time windows (by default 1 day) and prioritize the PRs based on whether or not a PR is likely to get a user update in the next time window, with the ones most likely to get an update considered as the most important. The approach was evaluated on the historical data of 475 projects, and gave the best performance using Random Forest model, with an accuracy of 0.85.

While the automatic prioritization is a useful approach for the
integrators, it offers little insight to the creators of the PR
about the quality of the submitted PR. More importantly,
PRioritizer sorts the PRs based on how likely they will have some user activity in the next time window, which may or may not indicate the PR will be merged, i.e. it is not necessarily indicative of the quality of the PR.
Keeping these limitations of the earlier approach in mind, our
\textbf{overarching Research Question was: What is the probability that
  a PR will be accepted within one month from the date of creation?}
We chose the time period of one month because only around 6.5\% of
the accepted PRs were accepted more than one month after creation, so
this time window is sufficient to approximate the probability that  a PR will ever be accepted. Moreover, a period of one month gives the project owners as well as the integrators sufficient time to plan which updates will likely be incorporated in near future.

In this study we want to investigate theoretical and practical
aspects of our overarching research question. We start from
an approximate replication of prior results by investigating how the
predictors listed in~\cite{prioritizer} work with this slightly
modified research question on a much larger dataset. Although this
is not an exact replication, the objective of our study is very
similar, and this attempt should put it more firmly in the context
of prior research.
Thus, the first goal of our study is:\\
\textbf{G1: How well does a model using the predictors listed
  in~\cite{prioritizer} perform in predicting PR's acceptance?}

Next, we wanted to understand if the model can be improved if we
take into account more comprehensive information about the PR
itself, the base project to which the PR is submitted, and the
characteristics of the PR creator (contributors' track record and experience, both of which are indicators of expertise,  were seen to have an effect of PR evaluation latency~\cite{yu2015wait} and PR acceptance~\cite{rahman2014insight}). 
These studies looked into the developers' expertise into the specific project where the PR is submitted. However, we hypothesize that the developers' activity, another indicator of expertise, across all OSS projects should also have an impact on the quality of PRs they submit. 
To test this hypothesis we obtained a much more comprehensive data set representing the
complete history of activities of the contributors and projects,
where we included author activities with all OSS projects, not just the
4218 projects under study (details in Sections~\ref{s:method}
and~\ref{s:data}). We hypothesize that variables related to the PR
creator's activity and historical performance, the properties of the
PR,  the characteristics of the base repository (to which the PR is being submitted to), and that of the head repository (the fork/clone from which the PR is being submitted) may have an impact on the probability of the PR being accepted.
Therefore, the second goal of our study is:\\
\textbf{G2: Can we more accurately predict the acceptance of a PR
  using a wide range of predictors characterizing the author, PR, and
  the base and forked repositories?}

While a well-performing model can predict whether a PR will be 
accepted or not with a high degree of confidence, it
does not provide insights to the PR authors who would like to
improve the chances of their PR being accepted. Specifically, we
wanted to illustrate how the probability of a PR of getting merged
may vary as the values the key predictors changes.
This was an insight requested by 86\% of the integrators the authors of~\cite{prioritizer} interviewed as well. So, the third goal of our study is:\\
\textbf{G3: How does the probability of a PR getting merged vary
  as the characteristics of the PR change?}

It is important to point out that the performance of a model in a
historic context may not represent the actual performance for
certain practical tasks. For example, as an integrator, we may want
to compare a PR opened a few weeks ago, that has an
extensive collection of comments, with a newly opened PR. 
Some PR characteristics, such as
the age of the PR or the number of comments it receives changes
after submission. The integrator would like to be able to prioritize PRs
immediately after submission, potentially reducing the time it takes
to accept important PRs. Another goal of our study is, therefore,
about predicting the probability of getting merged for the newly
submitted PRs so that the model would not penalize recently
submitted PRs.\\
\textbf{G4: Can we predict the probability of acceptance of a PR
  using information available at the time of PR submission?}

Finally, we want to illustrate how the model may be used in practice
and how well it would have performed for a specific large software
project. This is motivated by the fact that software projects change
over time and the model fitted on past data may not accurately
predict the acceptance of future PRs. Also, the information
available in practice at the time of PR submission would not include
some of the predictors for the best-performing model as discussed in
G4. The fifth goal of our study is:\\
\textbf{G5: How accurate would the PR acceptance model  be in a scenario
  representing actual use in a large project?}

To achieve these goals we obtained the necessary measurements and used a
tuned a Random Forest model. Our main findings are: (1) The model
using the predictors listed in~\cite{prioritizer} perform gave an
AUC-ROC value of 0.89 predicting PR acceptance rates. (2) Using
predictors characterizing the author, PR, and source and base
repositories we could achieve a much higher AUC-ROC value of
0.95. (3) We found the response curves to be non-monotonic for most
of the predictors (detailed result discussed in
Section~\ref{s:result}). (4) Excluding variables that change over the
lifetime of a PR resulted in an AUC-ROC value of 0.89. (5) Such predictor applied
to the NPM package
``bootstrap\footnote{https://www.npmjs.com/package/bootstrap}'', and
trained with only past data (prior to 2017-01-01) had AUC-ROC of
0.77 predicting acceptance of post 2017 PRs. If the model was
trained with a full set of 14 variables on the same past data, the AUC-ROC
increased to 0.94.

Our theoretical contributions include: (1) an approximate
replication study using the predictors in prior work to answer a
slightly different research question, and with a different and much
larger dataset; (2) identification of the different factors that are 
associated with PR acceptance (quality), and the response curves for
these predictors (how different values
of each of these predictors affect the chance of a PR getting
accepted); (3) result showing that the PR creator's activity in unrelated 
projects can also affect the quality of PRs they submit; and 
(4) identification of the factors that intuitively should affect the chance of
a PR getting accepted, but were found to be not necessary
to get an accurate predictions.

Our practical contributions
include: (1) a highly accurate predictive model relying on a set of
predictors that aren't difficult to calculate; (2) an approximate
``sweet spot'' in terms of the size of the PR that appears to significantly
increase the probability of acceptance of a PR (a potential recommendation
for the PR creators); (3) a well-performing model using only
predictors available at the time of PR creation; (4) an illustrative
example showing the application and performance of the model on a
large NPM package. 

Our data and code will be made available in our GitHub repository, which 
would enable other practitioners to use it for their research.

The rest of the paper is organized as follows: In
Section~\ref{s:relwork}, we discuss the related works in the
topic. In Section~\ref{s:method}, we discuss the Methodology. In
Section~\ref{s:data}, we describe the data collection and data
processing steps. In Section~\ref{s:result}, we describe the results
we found pertaining to our research goals. The implications of the
findings are discussed in Section~\ref{s:disc}. Finally, we discuss
the limitations of our study in Section~\ref{s:limit} and conclude
our paper in Section~\ref{s:conclusion}.

\vspace{-10pt}
\section{Related Work}\label{s:relwork}

Studies about Pull Requests are abundant both in terms of the number of studies and exploring the different topics associated to the PR based development. While majority of the works explored topics related to PR assignment, a number of studies, including a few case studies, explored the PR quality and PR acceptance scenario. 

The PRioritizer~\cite{prioritizer} study discusses the prioritization of the PRs for helping the integrators deal with multiple concurrent PRs. There are other studies that explore the perspective of the PR creators~\cite{gousios2016work} and the PR integrators~\cite{gousios2015work}, and list the challenges and practices in PR creation and merging scenario. These studies highlight that the primary challenges for the integrators are maintaining the project quality and dealing with multiple concurrent PRs, while the contributors face challenges regarding the compliance with the project guidelines for writing code and receiving timely feedback from the project owners. 

A number of studies describe various factors that influence the chance of PR getting accepted, like~\cite{soares2015acceptance}, which advocates using association rules to find the important factors, and found that the acceptance rates vary with the language the repository is written in, and also that having fewer commits, no additions, some deletions, some changed files, and the author having created a PR before and/or being part of the core team increase the chance of getting a PR accepted;~\cite{weissgerber2008small}, which indicates smaller PRs are more likely to get accepted;~\cite{yu2015wait}, which shows previously established track records of the contributors, availability and workload of the evaluators, and continuous integration based automated testing etc. have an impact on the latency of PR evaluation;~\cite{rahman2014insight}, which examined the effects of developer experience, language, calendar time etc. on the PR acceptance;~\cite{tsay2014influence}, which analyzed   the association of various technical and social measures with  the  likelihood  of  PR  acceptance.. 

There are a number of case studies that discuss the PR acceptance scenario in various OSS projects, like the Linux kernel~\cite{jiang2013will}, Firefox~\cite{baysal2012secret,mockus2002two}, Apache~\cite{mockus2002two} etc. 

There are other studies that described various aspects of the Pull Request based development model, like~\cite{gousios2014exploratory,GHPR,dabbish2012social}. A lot of studies about PRs  focused on finding the right evaluator for a particular PR, which is an important question as well, like~\cite{yu2014should,yu2014reviewer,jiang2017should,yu2016reviewer} to name a few. 

The NPM ecosystem is one of the most active and dynamic JavaScript ecosystems and~\cite{wittern2016look} presents its dependency structure and package popularity. Studies on NPM have mostly focused on its dependency networks~\cite{decan2018impact}, the popularity of NPM packages~\cite{dey2018software}, problems associated with library migration~\cite{zapata2018towards}, user participation patterns~\cite{dey2019patterns} etc.

\vspace{-10pt}
\section{Methodology}\label{s:method}

\begin{table*}[!htb]
\caption{List of the initial 50 variables }
\label{t:allvars}
\resizebox{\textwidth}{!}{%
\begin{tabular}{|
>{\columncolor[HTML]{FFFFC7}}p{4cm} 
>{\columncolor[HTML]{FFCCC9}}p{2cm} 
>{\columncolor[HTML]{FFCCC9}}p{4cm} 
>{\columncolor[HTML]{DAE8FC}}p{2.5cm} |
>{\columncolor[HTML]{EFEFEF}}p{2.5cm} |}
\hline
\multicolumn{1}{|p{4cm}|}{\cellcolor[HTML]{FFFFC7}\textbf{Total no. of PRs submitted by the creator before submitting this PR}} & \multicolumn{1}{p{2cm}|}{\cellcolor[HTML]{FFCCC9}\textbf{age of the PR}} & \multicolumn{1}{p{4cm}|}{\cellcolor[HTML]{9AFF99}\textbf{No. of PRs submitted to the repository previously}} & If the head repo is a fork & If the base repo is a fork \\\hline
\multicolumn{1}{|p{4cm}|}{\cellcolor[HTML]{FFFFC7}\textbf{Fraction of the total PRs submitted by the creator that are merged}} & \multicolumn{1}{p{2cm}|}{\cellcolor[HTML]{FFCCC9}Calendar date of PR creation} & \multicolumn{1}{p{4cm}|}{\cellcolor[HTML]{9AFF99}\textbf{Fraction of PRs submitted to the repository that are accepted}} & size of the head repo & size of the base repo \\\hline
\multicolumn{1}{|p{4cm}|}{\cellcolor[HTML]{FFFFC7}The creator's association with repository} & \multicolumn{1}{p{2cm}|}{\cellcolor[HTML]{FFCCC9}If the PR is marked as Closed} & \multicolumn{1}{p{4cm}|}{\cellcolor[HTML]{9AFF99}If the package associated is a direct dependency of any of the PR creator's repositories} & no. of stars of the head repo & no. of stars of the base repo \\\hline
\multicolumn{1}{|p{4cm}|}{\cellcolor[HTML]{FFFFC7}If the creator submitted PRs to this repository previously} & \multicolumn{1}{p{2cm}|}{\cellcolor[HTML]{FFCCC9}\textbf{No. of discussion comments}} & \multicolumn{1}{p{4cm}|}{\cellcolor[HTML]{9AFF99}No. of NPM packages associated with the repository} & If the head repo had issues enabled & If the base repo had issues enabled \\\hline \cline{3-3}
\multicolumn{1}{|p{4cm}|}{\cellcolor[HTML]{FFFFC7}If the user had any PR accepted in this repository} & \textbf{No. of code review comments} & \multicolumn{1}{p{4cm}|}{\cellcolor[HTML]{FFCCC9}If the PR is from a different branch of the same repository} & If the head repo had wiki enabled & If the base repo had wiki enabled \\\hline
\multicolumn{1}{|p{4cm}|}{\cellcolor[HTML]{FFFFC7}If the PR creator was the integrator for some other PR previously} & \textbf{No. of lines added} & \multicolumn{1}{p{4cm}|}{\cellcolor[HTML]{FFCCC9}If the PR contains fix for some issue} & If the head repo had projects enabled & If the base repo had projects enabled \\\hline
\multicolumn{1}{|p{4cm}|}{\cellcolor[HTML]{FFFFC7}If this is the first time the creator is submitting a PR} & \textbf{No. of lines deleted} & \multicolumn{1}{p{4cm}|}{\cellcolor[HTML]{FFCCC9}If the PR contains any test code} & If downloads were enabled for the head repo & If downloads were enabled for the base repo \\\hline
\multicolumn{1}{|p{4cm}|}{\cellcolor[HTML]{FFFFC7}\textbf{Total no. of commits by the PR creator across all Git projects}} & \textbf{No. of files touched} & \multicolumn{1}{p{4cm}|}{\cellcolor[HTML]{FFCCC9}If the PR description has any HTML tag in it} & No. of forks of the head repo & No. of forks of the base repo \\\hline
\multicolumn{1}{|p{4cm}|}{\cellcolor[HTML]{FFFFC7} \textbf{Total no. of blobs touched by the PR creator across all Git projects}} & \textbf{No. of commits in the PR} & \multicolumn{1}{p{4cm}|}{\cellcolor[HTML]{FFCCC9}If the PR description has any emoji in it} & No. of issues for the head repo & No. of issues for the base repo \\\hline \cline{2-3}
\textbf{Total no. of Git projects the PR creator contributed to} & \multicolumn{1}{p{2cm}|}{\cellcolor[HTML]{FFFFC7} Fraction of creator's commits in the repo} & \cellcolor[HTML]{DAE8FC}If the head repo was deleted after the PR was submitted & Calendar date of head repo creation & Calendar date of base repo creation \\ \hline
\end{tabular}%
}
\vspace{-10pt}
\end{table*}

\begin{figure}[!t]
\centering
\vspace{-10pt}
\includegraphics[width=\linewidth]{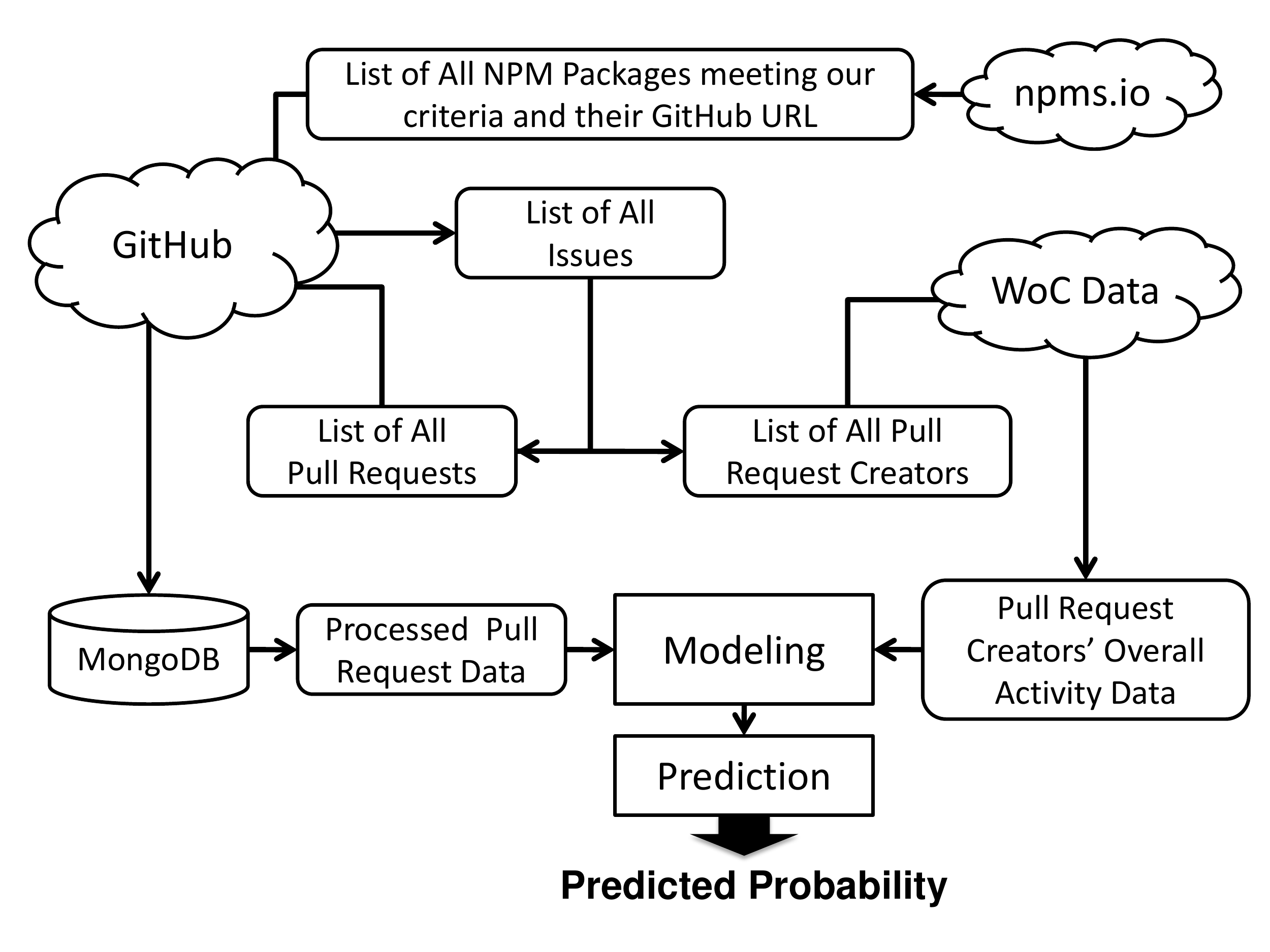}%
\vspace{-10pt}
\caption{The Data Collection and Modeling Architecture.}
\label{fig:flow}
\vspace{-20pt}
\end{figure}

\subsection{Data Collection}
To answer the Research Question and address the different goals listed in Section~\ref{s:intro}, we first needed the list of all NPM packages that satisfy our criteria of having more than 10,000 downloads per month and a GitHub repository with at least one PR. This was obtained from the \texttt{npms.io} website, using the API provided~\footnote{https://api.npms.io/v2/package/[package-name]}. The associated GitHub repository URLs were collected from their metadata information, which was obtained by using a ``follower" script, as described in NPM's GitHub repository~\footnote{https://github.com/npm/registry/blob/master/docs/follower.md}. After filtering for our criteria that the NPM package must have more than 10,000 monthly downloads (since January, 2018), a functional link to its GitHub repository, and at least one PR, we were left with 4218 different NPM packages.

Next, we needed the list of all PRs for these NPM packages. To obtain this, we first collected all the issues associated with these NPM packages, since GitHub considers PRs as issues, and then identified the issues that have an associated patch, i.e. the ones that are Pull Requests. The list of all issues for the packages was obtained using the GitHub API for issues\footnote{https://developer.github.com/v3/issues/}, using the \texttt{state=all} flag. 
Out of all the issues, we identified 483,988 PRs for the 4218 packages (until January, 2019, when the data was collected).
It is worth mentioning here that sometimes more than one NPM package can have the same associated GitHub repository, e.g. all TypeScript NPM packages (starting with ``@types/'', like @types/jasmine, @types/q, @types/selenium-webdriver etc.) refer to GitHub repository\\ ``DefinitelyTyped/DefinitelyTyped''. To avoid double-counting and further confusion, we saved the issues keying on the repository instead of the package name, though we also saved the list of packages associated with a repository. We found that there are 3601 unique repositories associated with these 4218 packages.

Then we obtained the data on all the PRs from GitHub using the API\footnote{https://developer.github.com/v3/pulls/}. This data was stored in a local MongoDB database. We used a Python script to extract the data from this database and process it into a CSV file that was used for modeling. More details about the data are discussed in Section~\ref{s:data}.

The data on the PR creators' overall activity across all projects that use Git were obtained from a recent version (version \textbf{V}) of the WoC (World of Code) data~\cite{woc19}. WoC is a prototype of an updatable and expandable infrastructure to support research and tools that rely on
version control data from the entirety of open source projects that use Git. 
Specifically, we used it to compile the profiles of PR authors. We
identified PR authors by obtaining commits they included in their
PRs. We then identified the commit authors for these commits in WoC
using commit to author maps. Then we identified all the remaining
commits for these authors using author to commit map. That full set
of commits for each author was used to count projects (via commit
to project map), files, and blobs (specific versions of a source
code file). We only counted blobs created by
the author (by verifying that the author mad the first commit
creating the said blob) to reduce the noise produced when developers clone
repositories, thus creating a massive number of blobs that were 
authored by others. To construct relevant measures for the PR
prediction we only used the commits made by the PR author prior to
the creation of the PR, thus reconstructing the state of affairs as
it existed at the time of PR creation. 


The data from these two sources (the PR data from GitHub and the  PR creators' activity data collected using the WoC tool) were consolidated to \textit{construct} the final dataset we used for modeling. We sorted the data by the PR creation date, and extracted relevant variables from it, making sure only to use historical information while constructing different variables. More details on the data is discussed in Section~\ref{s:data}.

\vspace{-10pt}
\subsection{Data Analysis}

We performed our data analysis (modeling) tasks in R. Our response variable for modeling was the binary variable indicating whether a PR is merged with one month (30 days) after the date of creation or not. 
We looked into several modeling approaches for predicting the probability of acceptance of the PRs: Logistic Regression, Generalized Additive Models, Support Vector Machines, Random Forest, and Multi-Layer Perceptron networks. We tested all of these approaches on our dataset with the full set of predictors (listed in Section~\ref{s:data}), and found that a tuned Random Forest model performs better than the others. Therefore, we decided to use it as our modeling methodology for this study. 

To obtain the optimal number of predictors we used the ``rfcv'' function from the \textit{randomForest} R package, which shows the cross-validated prediction performance of models with sequentially reduced number of predictors (ranked by variable importance) via a nested cross-validation procedure. 

We used the ``train'' function from the \textit{caret} package in R for performing a grid search (using a 10 fold cross-validation) on the training data to find the optimal values of the model parameters: ``ntree" (number of trees to grow) and ``mtry" (number of variables randomly sampled as candidates at each split), that gives the highest Accuracy.
\textit{The optimum value for ``ntree'' was found to be 500, and the optimum value of ``mtry'' was found to be 6.}

To calculate the performance of the final tuned model, we decided to use 70\% of the data, selected randomly, as our training set, and the other 30\% as our test set.
Models with this configuration were used to predict the final probabilities of the PRs getting merged under different conditions pertaining to the different research goals we have.

We wanted to test the viability of our approach and the predictive performance of our model in a realistic scenario. So we decided to look into one NPM package as an example, and see if historical information about the package can be used to predict whether a PR submitted in future will be accepted or not. We needed a package that has sufficient number of PRs, so that we have enough historical data to train the model and enough PRs in  the test set, so that the test results are trustworthy.  We selected the ``bootstrap''\footnote{https://www.npmjs.com/package/bootstrap} NPM package as an example since it is a popular package that matches our criteria. The details of the data on this package are discussed in Section~\ref{s:data}.

\begin{table*}[!htb]
\caption{Descriptive Statistics and Detailed definition of the selected variables}
\label{t:vars}
\resizebox{\textwidth}{!}{%
\begin{tabular}{lllllll}
\hline
\textbf{Variable Name} & \textbf{Variable Description} & \textbf{5\%} & \textbf{Median} & \textbf{Mean} & \textbf{95\%} & \textbf{Distribution Plot} \\\hline
age & Seconds between PR creation and PR closure (max 30 days = 30*24*3600) & 231 & 100*1e3 & 651*1e3 & 2.5*1e6 & \includegraphics[width = 2cm, height = 0.4 cm]{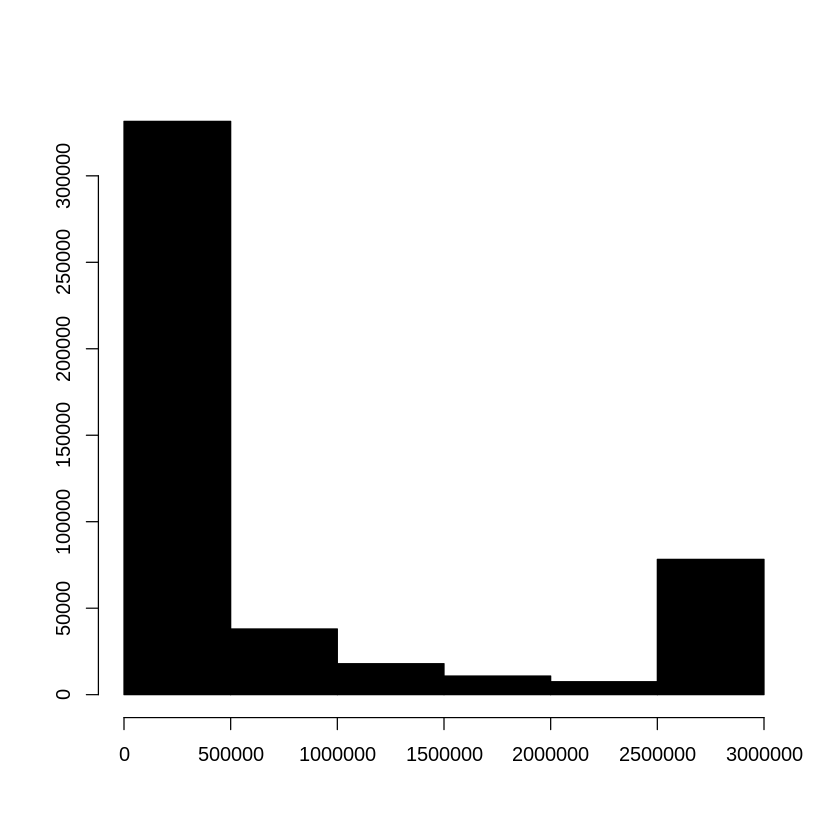} \\
commits & Number of commits in the Pull Request & 1 & 1 & 4 & 7 &   \includegraphics[width = 2cm, height = 0.4 cm]{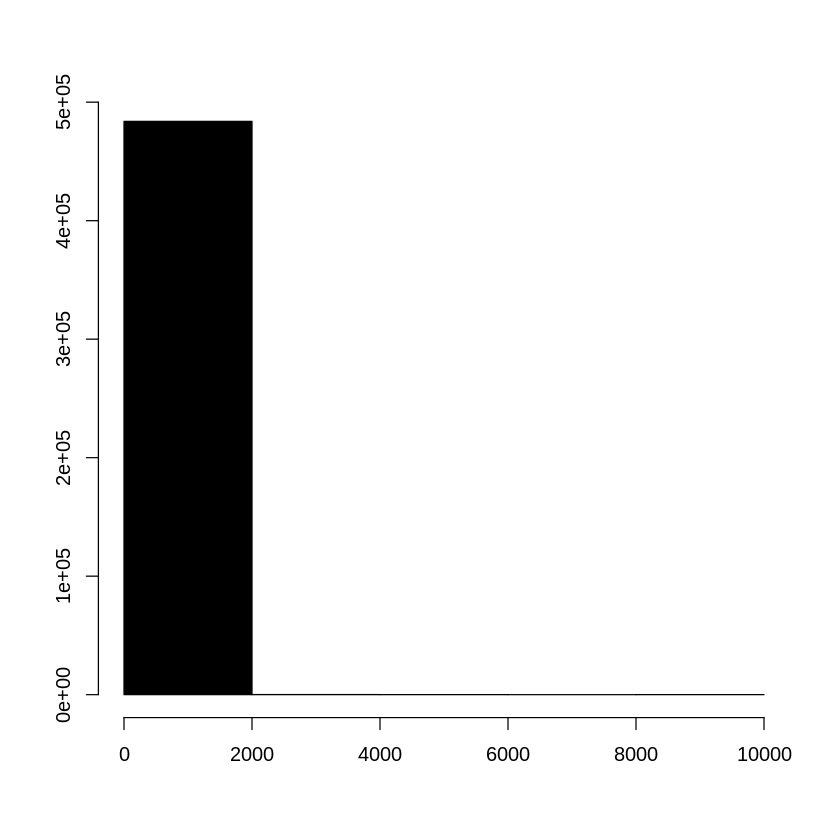}\\
changed\_files & Number of files modified in the Pull Request & 1 & 2 & 10 & 17 & \includegraphics[width = 2cm, height = 0.4 cm]{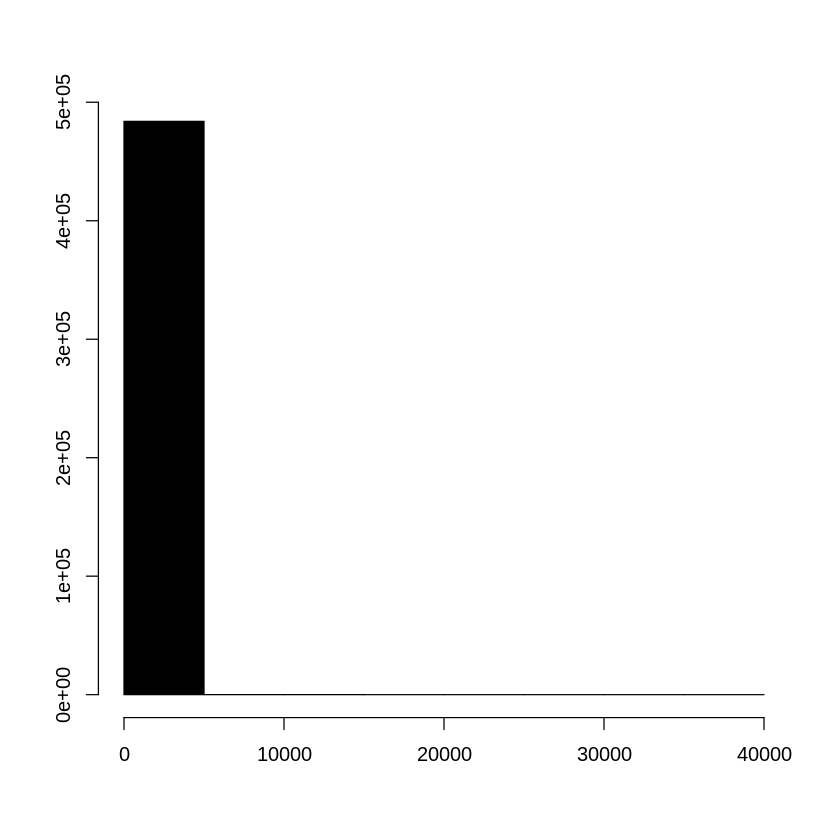} \\
comments & Number of discussion comments against the Pull Request & 0 & 2 & 3 & 11 & \includegraphics[width = 2cm, height = 0.4 cm]{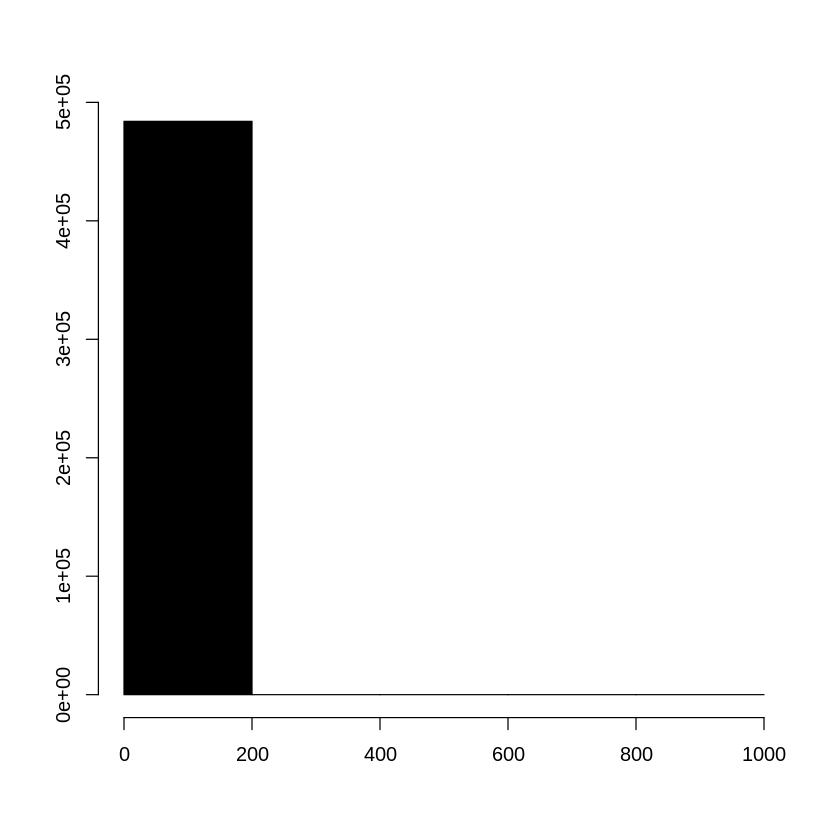} \\
review\_comments & Number of code review comments against the Pull Request & 0 & 0 & 1 & 6 &  \includegraphics[width = 2cm, height = 0.4 cm]{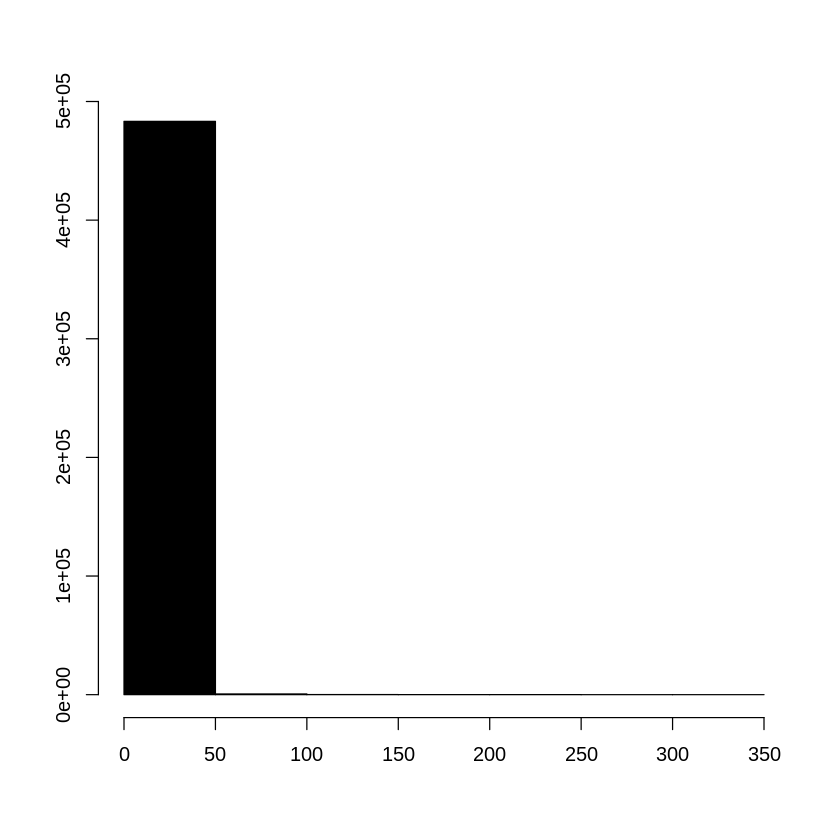}\\
additions & Number of lines added in the Pull Request & 1 & 12 & 703 & 619 &  \includegraphics[width = 2cm, height = 0.4 cm]{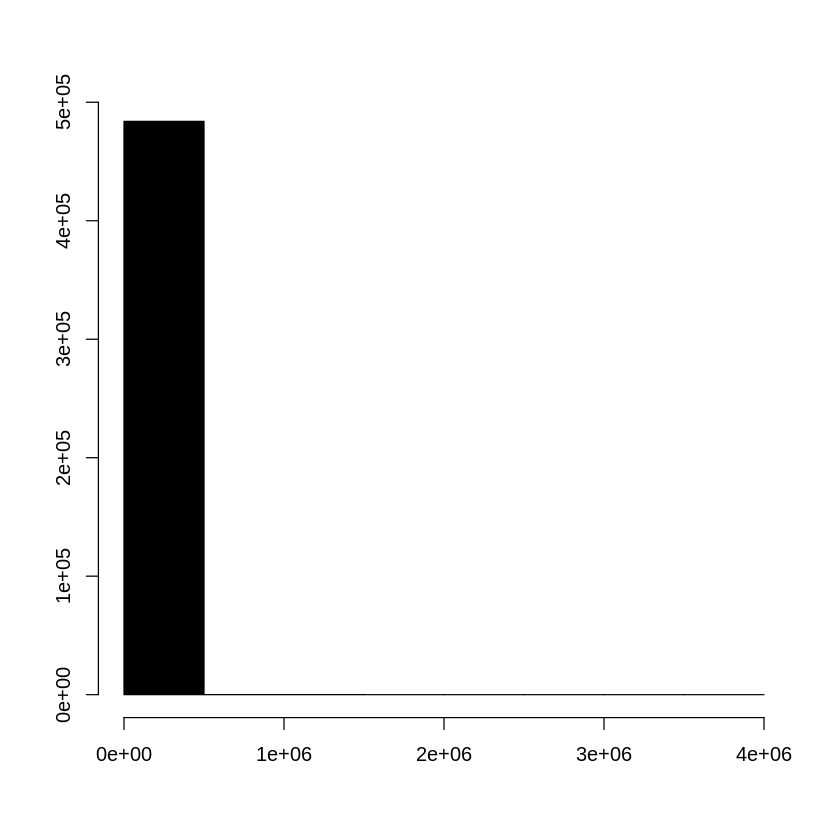}\\
deletions & Number of lines deleted in the Pull Request & 0 & 2 & 385 & 248 & \includegraphics[width = 2cm, height = 0.4 cm]{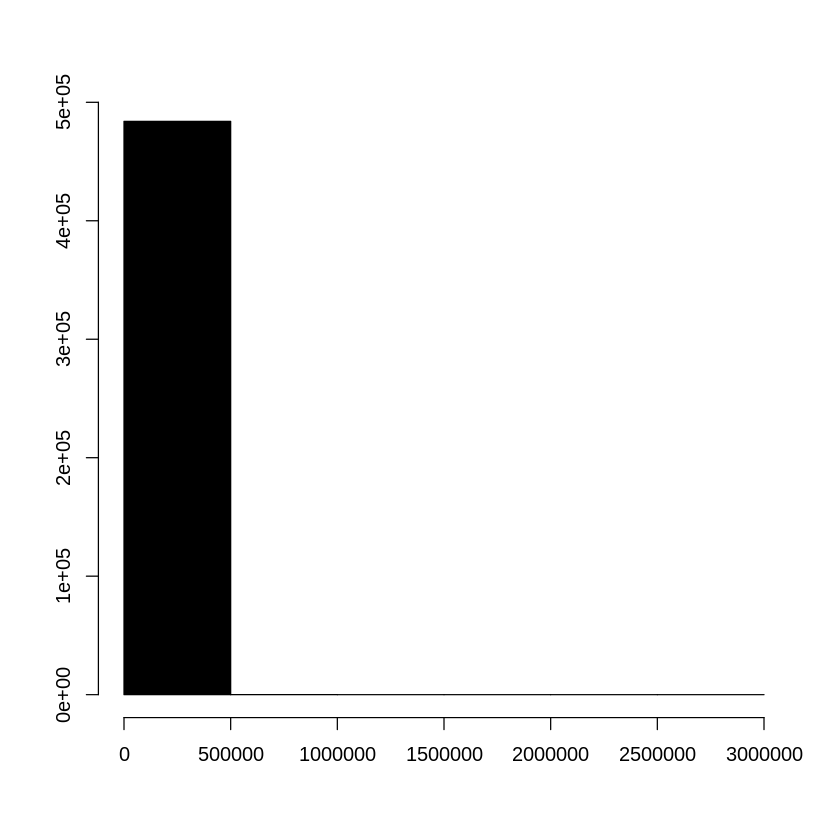}  \\
creator\_total\_commits & Total number of commits made by the PR creator across Git Projects & 4 & 786 & 9847 & 12,386 & \includegraphics[width = 2cm, height = 0.4 cm]{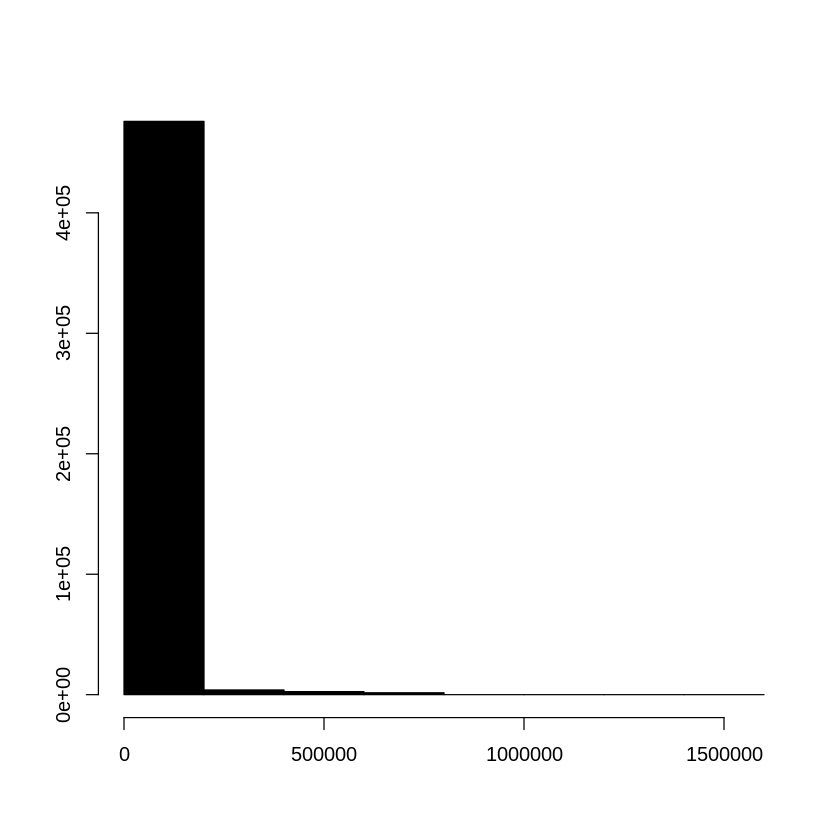} \\
creator\_total\_blobs & Total number of blobs authored by the PR creator across Git Projects & 6 & 2080 & 12,308 & 40,808 & \includegraphics[width = 2cm, height = 0.4 cm]{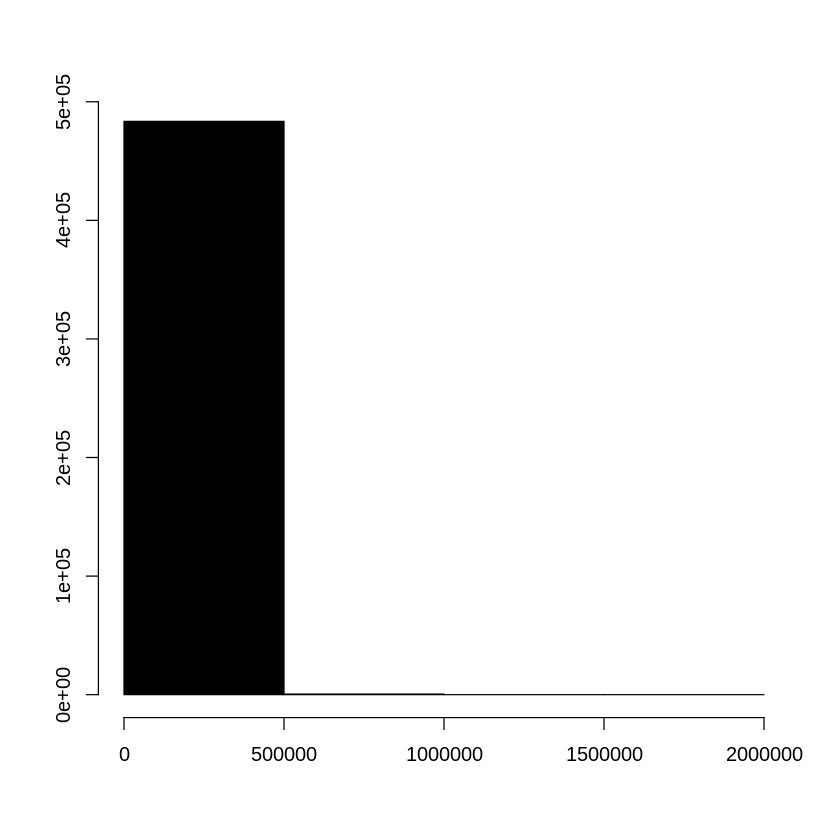} \\
creator\_total\_projects & Total number of projects the PR creator contributed to across Git Projects & 3 & 1632 & 6481 & 31,880 &  \includegraphics[width = 2cm, height = 0.4 cm]{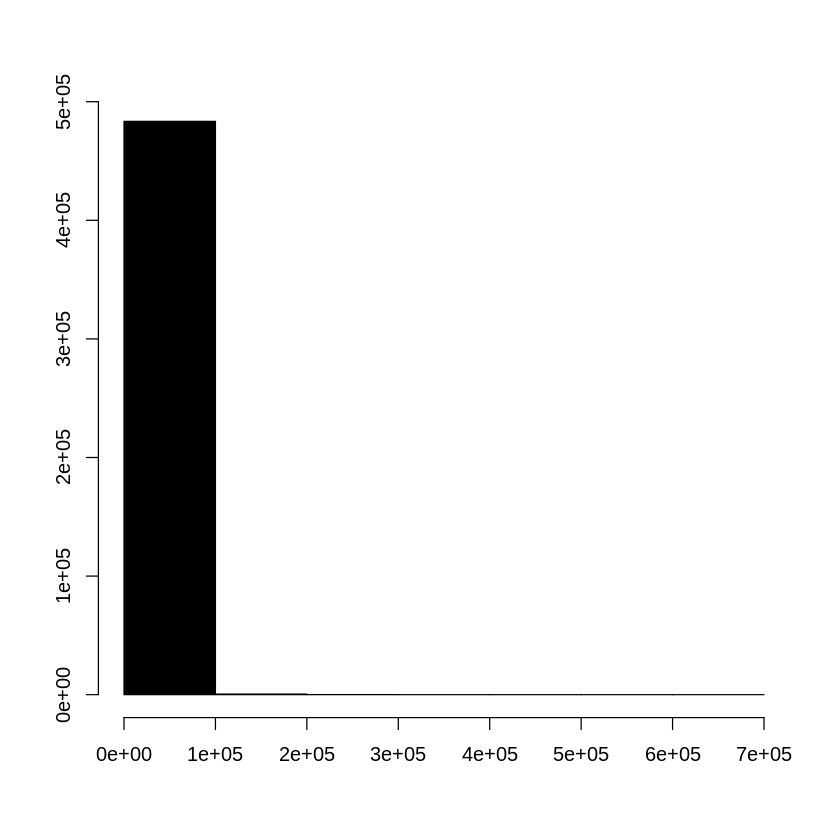}\\
repo\_submitted & Number of PRs submitted against the repository  & 9 & 787 & 4787 & 30,270 & \includegraphics[width = 2cm, height = 0.4 cm]{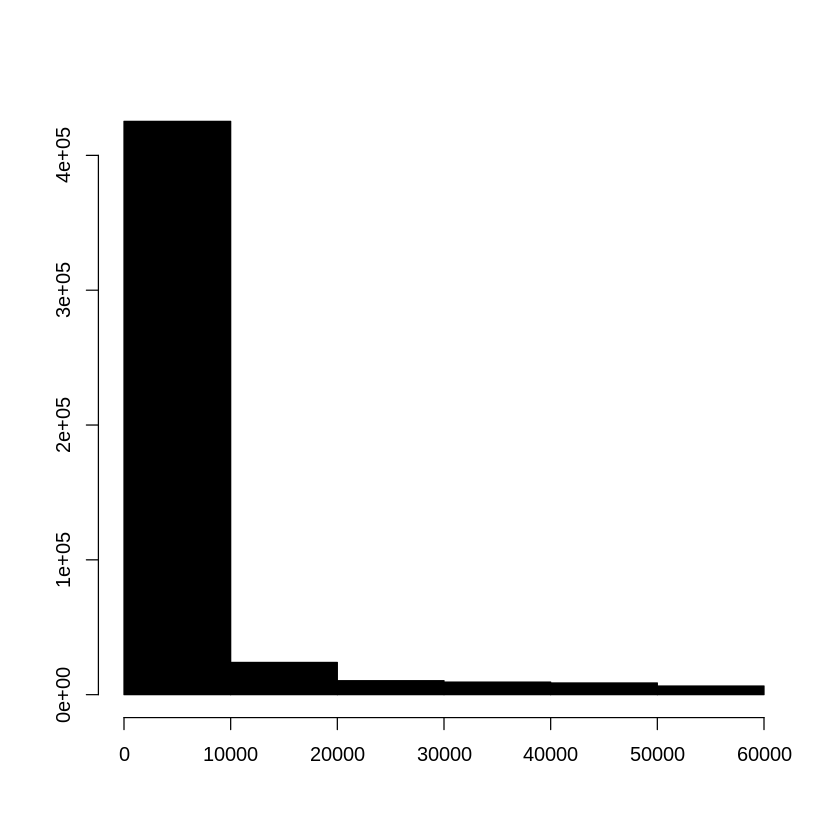} \\
repo\_accepted & Fraction of the submitted  PRs accepted by the repository & 0.1 & 0.70 & 0.63 & 0.91 &  \includegraphics[width = 2cm, height = 0.4 cm]{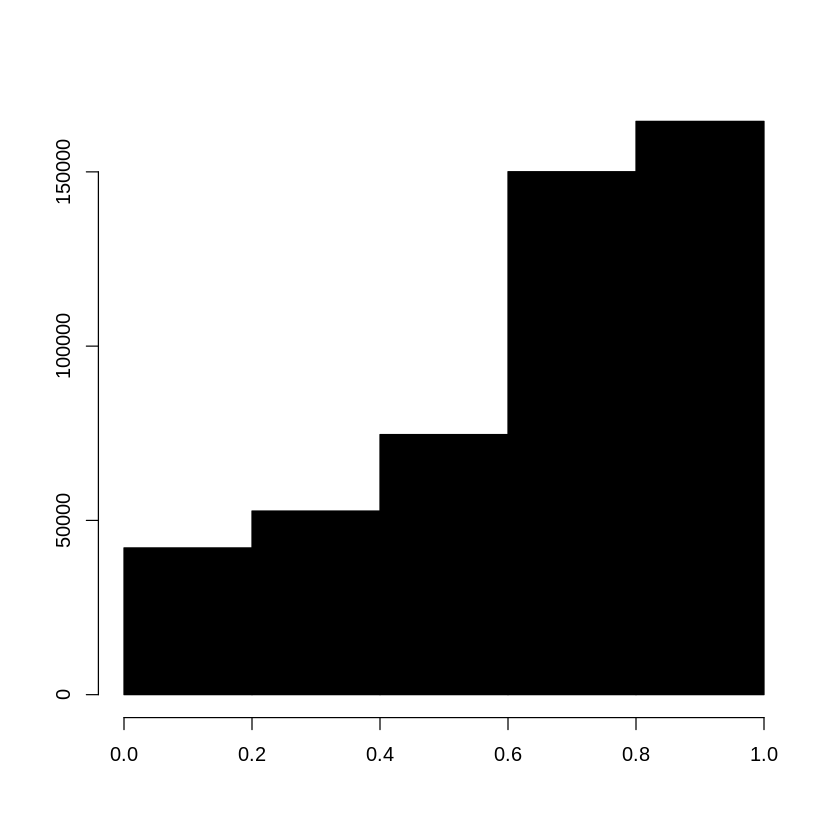}\\
creator\_submitted & Number of PRs submitted by the PR creator across NPM projects & 0 & 12 & 282 & 1043 &  \includegraphics[width = 2cm, height = 0.4 cm]{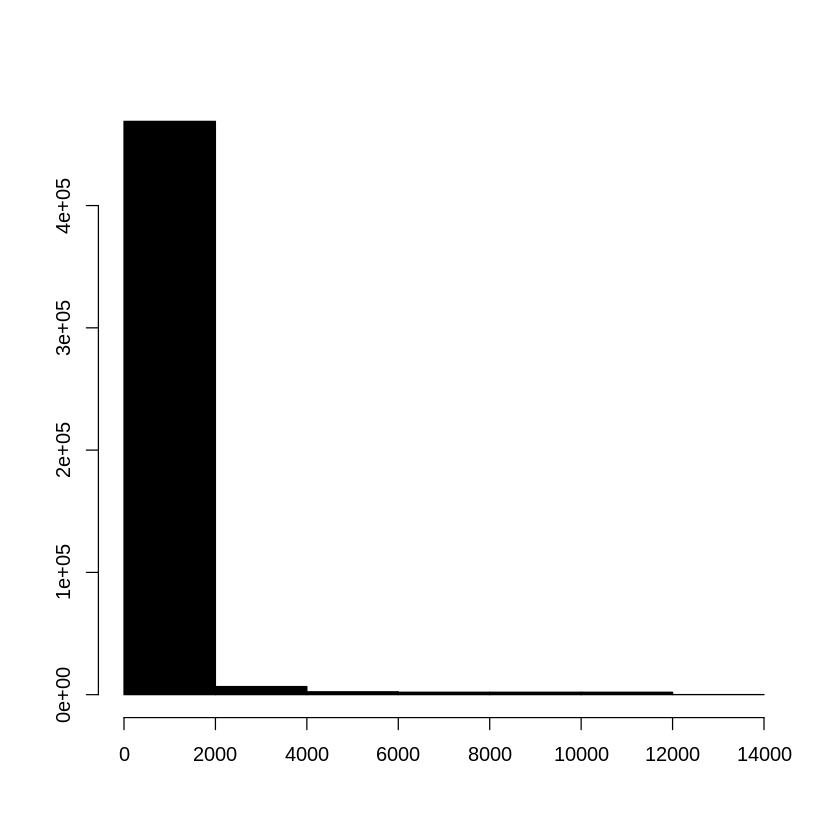}\\
creator\_accepted & Fraction of PRs submitted by the PR creator accepted across NPM projects & 0 & 0.64 & 0.53 & 1.00 &  \includegraphics[width = 2cm, height = 0.4 cm]{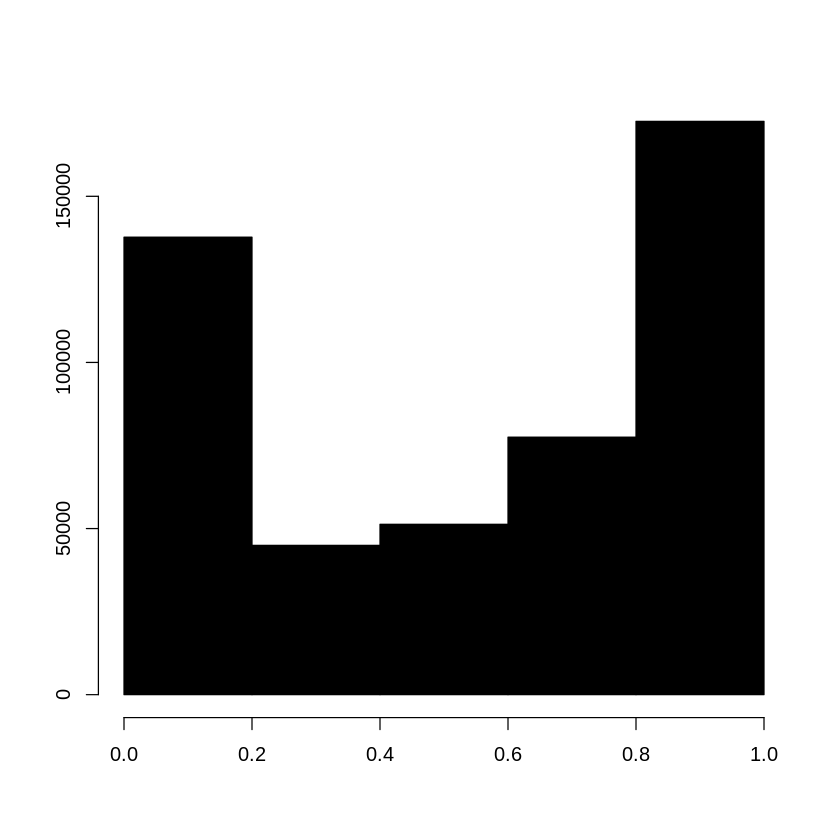}\\\hline
\end{tabular}%
}
\vspace{-10pt}
\end{table*}

\vspace{-10pt}
\section{Data Description}\label{s:data}

In this section, we present the original set of variables we started with, and the final set of variables we used for our final models. 

We were careful to include the variables listed in~\cite{prioritizer} in our dataset, however, we later decided not to use the variable: ``Last Comment Mention", since (1) while it might be important to predict whether a PR will see some activity in the next time window (which was the goal of~\cite{prioritizer}), it is unlikely to have as much importance in predicting whether a PR will be accepted or not, and more importantly, (2) for a number of PRs, there are comments on it even after it is merged\footnote{e.g. https://github.com/twbs/bootstrap/pull/29257}, and given~\cite{prioritizer} only dealt with open PRs, it is hard to replicate the variable in our context. 

We extracted 50 variables from the PR data collected using the GitHub API and the PR creators' activity data collected using the WoC tool. The list of the description of these variables is presented in Table~\ref{t:allvars}. The variables can broadly be grouped into 5 categories, that describe different characteristics of : (1) The PR creator (the yellow box), (2) The specific PR (red box), (3) The NPM package repository under consideration (green box), (4) The specific characteristics of the head repository, from which the PR is submitted (blue box), and (5) The specific characteristics of the base repository to which the PR is submitted (gray box).

Using the ``rfcv'' function, as mentioned in Section~\ref{s:method}, we found that 14 variables give the optimum result. So, first we created a Random Forest model using all the variables as predictors, and then we selected the top 14 predictors by looking at the variable importance. The variables that were selected in our final set of predictors are marked in bold in Table~\ref{t:allvars}. The variable names used in the models, along with their descriptions, descriptive statistics, and distribution plots are shown in Table~\ref{t:vars}. The variable \textit{age} was measured in seconds. Since our Research Question was about finding out if a PR is merged within one month of creation, we used 30 days, or $30*24*3600 = 2,592,000$ seconds as the max value for \textit{age}. 

The values of \textit{creator\_submitted} and \textit{repo\_submitted} for a particular PR are the number of PRs that were submitted, before that particular PR was submitted. The fractions representing PR acceptance by the repository (\textit{repo\_accepted}) and for the PR creator (\textit{creator\_accepted}) are also measured by only considering the PRs that were submitted and accepted before that particular PR was submitted. Therefore the values of these four variables varied for different PRs.

Since the original distributions for most of the variables were very skewed, we decided to log transform the data before using them in the models. 

It is worth mentioning that the PR creator's association with the
repository where the PR was submitted was one of the most important
predictors according to variable importance, similar to what was
observed by~\cite{tsay2014influence}. However, we decided not to use
this variable in our final dataset since the value of this field can
be updated retroactively (e.g. a PR creator, who had no association
with a project when first submitting a PR, might become
a \textit{member} later, and the corresponding field in the first PR
might be updated after its acceptance), and we have no way to know the creator's
association at the time a PR was submitted, or to verify that the affiliation wasn't updated retroactively (see Section~\ref{s:disc} for examples to the contrary), which would be required
to faithfully reconstruct the data as it were at the time the PR was
created. Instead of using this variable, we created two binary
variables indicating whether the creator created any PR to the
particular repository before and if any of those PRs had been
accepted, but both of them proved to be not very significant. We,
therefore, suspect that its importance in prior work could be due to the so
called data leakage~\cite{tu2018careful}, when 
the information leaked from the ``future'' makes prediction models misleadingly optimistic.

As mentioned in Section~\ref{s:method}, we selected the NPM package ``bootstrap'' to test the viability of our model in a practical scenario. The package had 8425 PRs submitted against it, out of which 4589 (54.5\%) were merged. We decided to look at the PRs submitted until 2016-12-31 as our training data, and PRs submitted on or after 2017-01-01 as our test data, which left 6436 (76\% of the PRs for this package) PRs in the training set. However, the two sets weren't equally balanced, since 42\% of the PRs in the training set were accepted, but 58\% of the PRs in the test set were accepted. Still, we went ahead with using this dataset to address our Goal 5.

\vspace{-10pt}
\section{Results}\label{s:result}

In this section we discuss our findings and address the Research Goals listed in Section~\ref{s:intro}, and present some general statistics about the data.

\vspace{-10pt}
\subsection{General Statistics about the Data}\label{ss:gen}
Here we discuss some general statistics, which, in spite of not being directly related to our research goals, can give us some insight into the data and the NPM ecosystem in general. 

To recap, our study focused on 4218 NPM packages (3601 unique GitHub repositories) with more than 10,000 monthly downloads since January, 2018, an active GitHub repository, and at least one PR. We collected 483,988 pull-requests, which were created by 82,142 creators.

A few interesting statistics about the data are reported below:
\begin{itemize}
    \item 291,089 (60\%) of the total PRs that were submitted were merged (accepted).
    \item We found that 39,570 (48.2\%) of the creators created only one PR (one time contributors), and only 6523 (7.9\%) creators created 10 or more PRs.
    \item 294 (8.2\%) repositories received only one PR, and 535 (14.9\%) of the repos received 100 or more PRs.
    \item 397583 (82\%)  PRs were created for NPM packages that any project the PR creator contributed to depend on directly.
    \item only 11357 (2.3\%) PRs were still open when the data was collected.
    \item 312409 (65\%) head repositories from which PRs were submitted were forks of the original (base) repositories.
    \item 209216 (43\%) of the PRs submitted contained fixes for some bugs.
\end{itemize}

\begin{figure}[!t]
\centering
\includegraphics[width=0.65\linewidth]{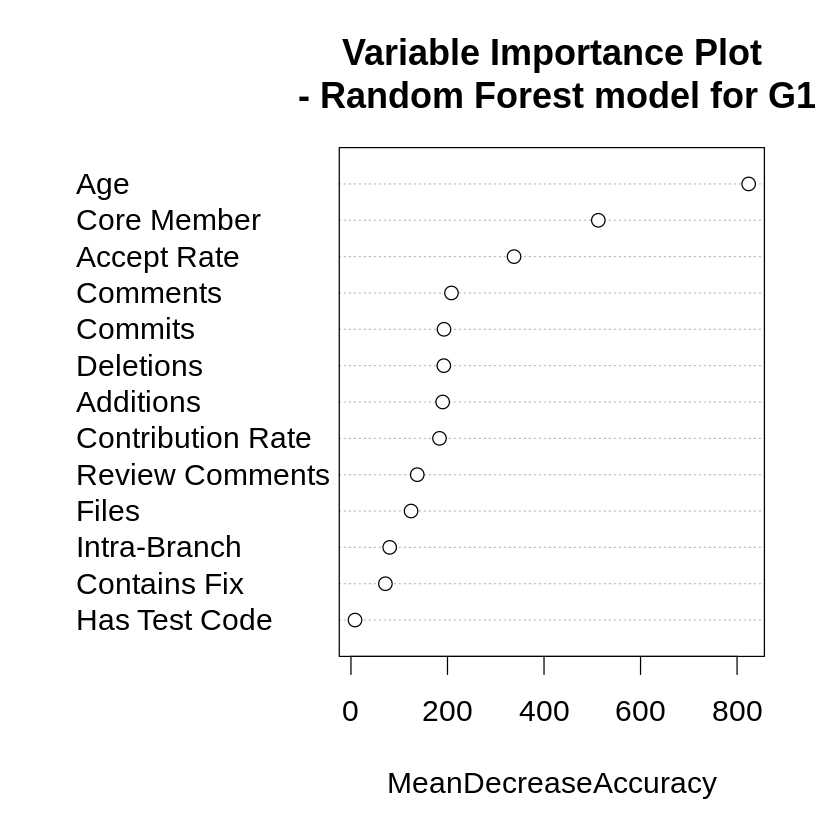}%
\vspace{-10pt}
\caption{Variable Importance plot for the Random Forest model using predictors used in~\cite{prioritizer}}
\label{fig:rf1}
\vspace{-10pt}
\end{figure}

\begin{figure}[!t]
\centering
\vspace{-10pt}
\includegraphics[width=0.65\linewidth]{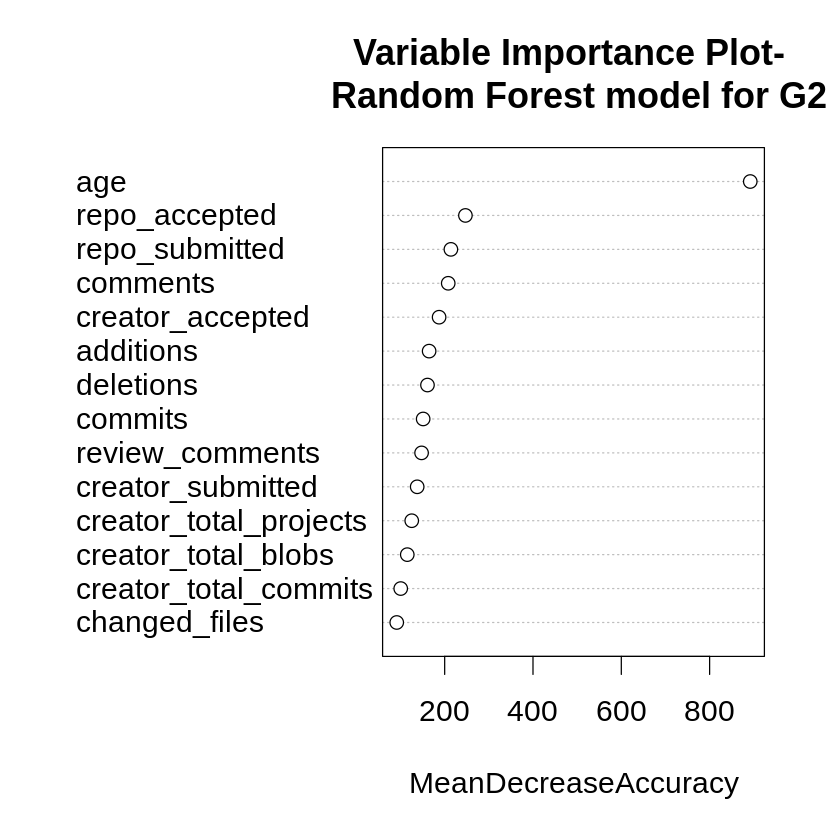}%
\vspace{-10pt}
\caption{Variable Importance plot for the Random Forest model using our set of 14 predictors}
\label{fig:rf2}
\end{figure}

\begin{figure}[!t]
\centering
\vspace{-10pt}
\includegraphics[width=0.65\linewidth]{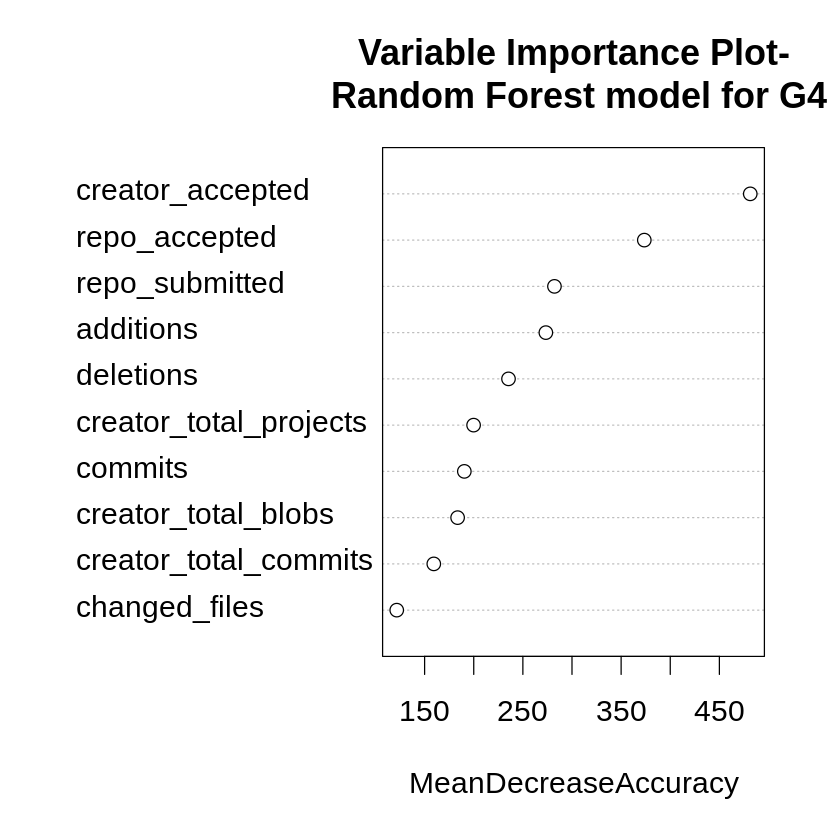}%
\vspace{-10pt}
\caption{Variable Importance plot for the Random Forest model without using the time dependent predictors}
\label{fig:rf4}
\end{figure}

\begin{figure}[!t]
\centering
\vspace{-10pt}
\includegraphics[width=0.65\linewidth]{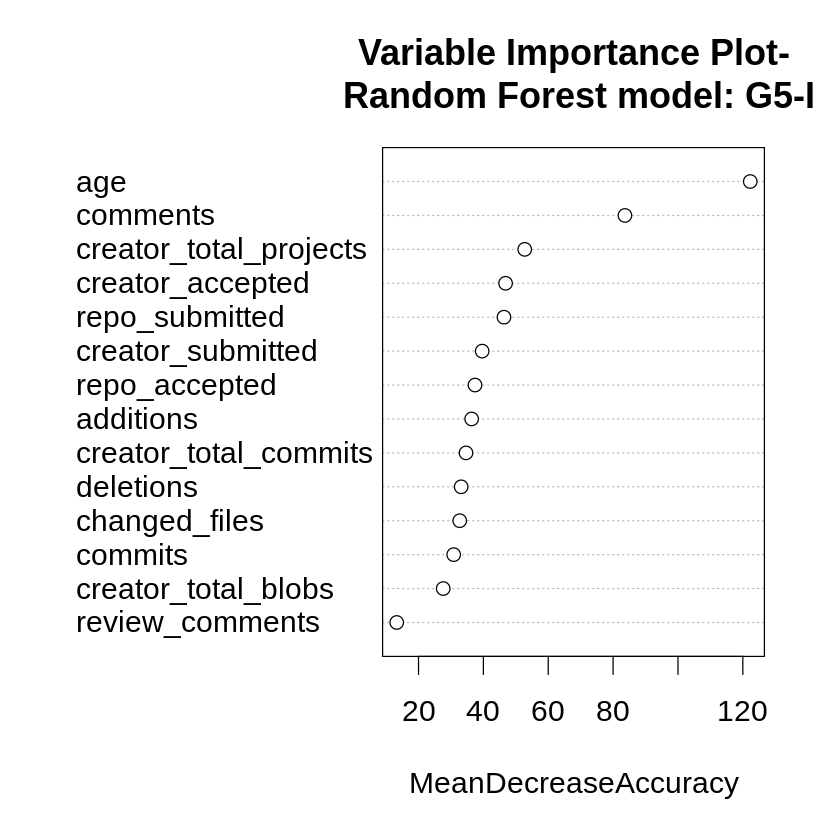}%
\vspace{-10pt}
\caption{Variable Importance plot for the Random Forest model for NPM package ``bootstrap'' using all predictors}
\label{fig:rf51}
\vspace{-10pt}
\end{figure}

\begin{figure}[!t]
\centering
\vspace{-10pt}
\includegraphics[width=0.65\linewidth]{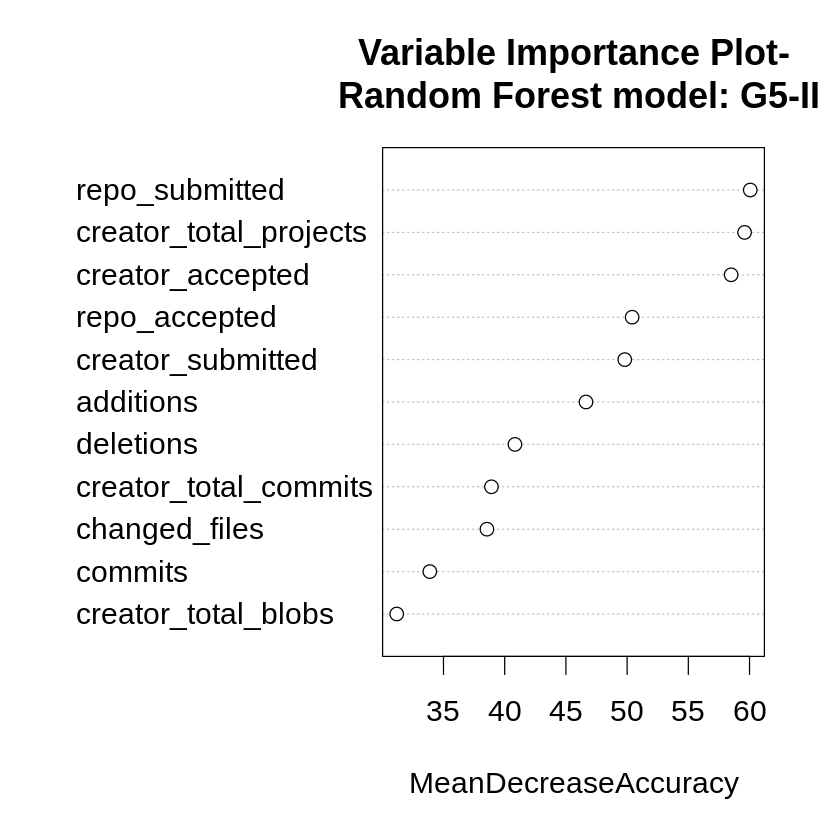}%
\vspace{-10pt}
\caption{Variable Importance plot for the Random Forest model for NPM package ``bootstrap'' without using the time dependent predictors}
\label{fig:rf52}
\vspace{-10pt}
\end{figure}

\begin{table}[!htb]
\caption{Partial Dependence plots for the predictors according to the Random Forest model created for G2, showing how different values of the different variables affect a PR's chance of being accepted}
\label{t:pplot}
\resizebox{\linewidth}{!}{%
\begin{tabular}{ll}
\includegraphics[width = 0.3\linewidth, height = 0.08\textheight]{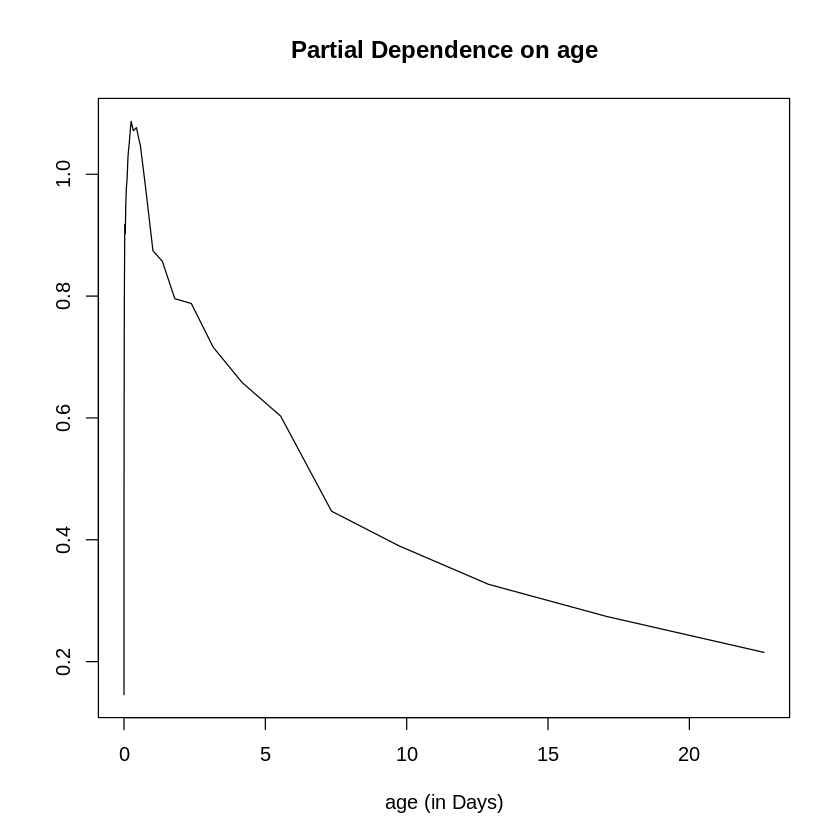} &
\includegraphics[width = 0.3\linewidth, height = 0.08\textheight]{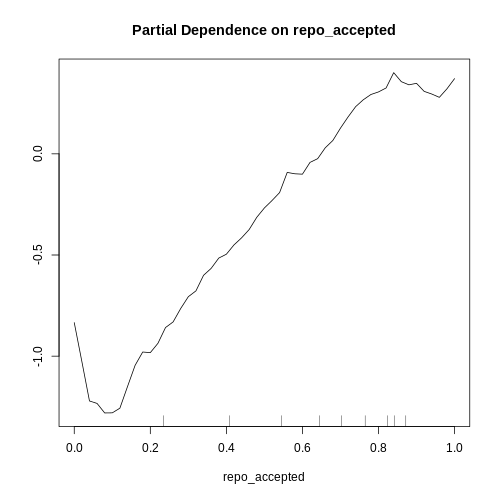} \\
\includegraphics[width = 0.3\linewidth, height = 0.08\textheight]{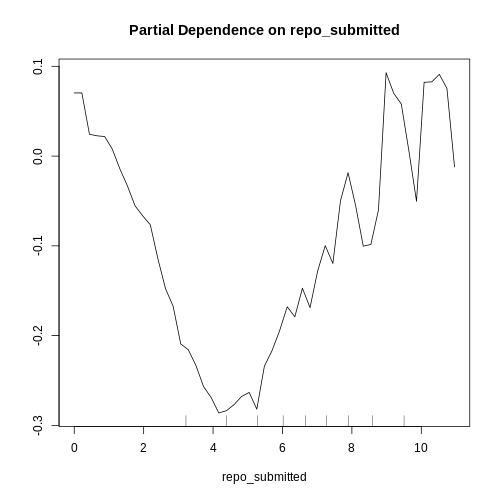} &
\includegraphics[width = 0.3\linewidth, height = 0.08\textheight]{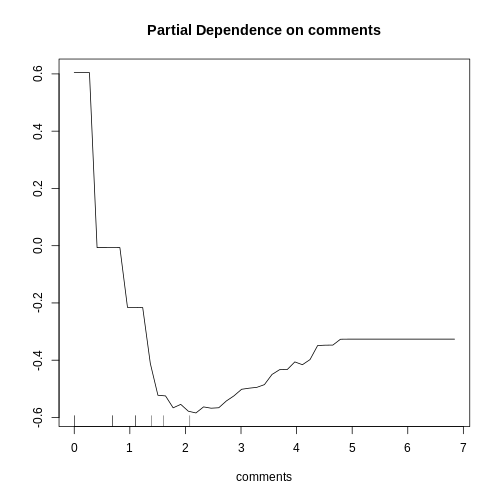} \\ 
\includegraphics[width = 0.3\linewidth, height = 0.08\textheight]{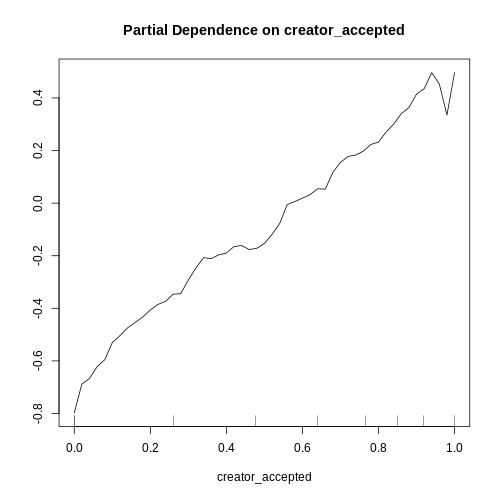} &
\includegraphics[width = 0.3\linewidth, height = 0.08\textheight]{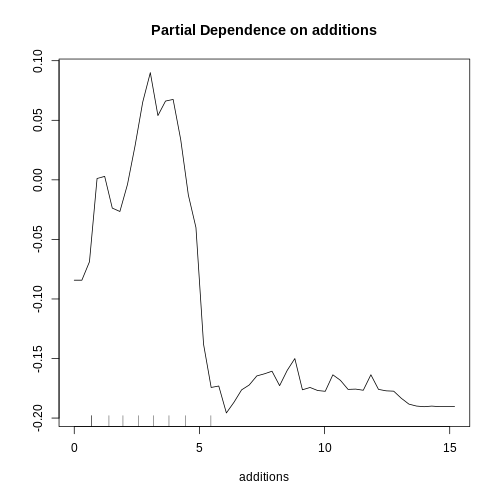} \\
\includegraphics[width = 0.3\linewidth, height = 0.08\textheight]{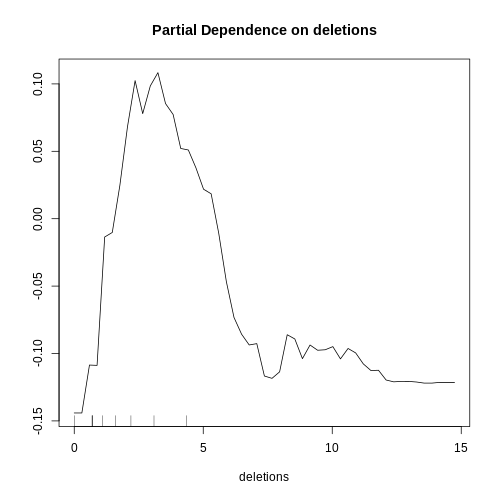} & \includegraphics[width = 0.3\linewidth, height = 0.08\textheight]{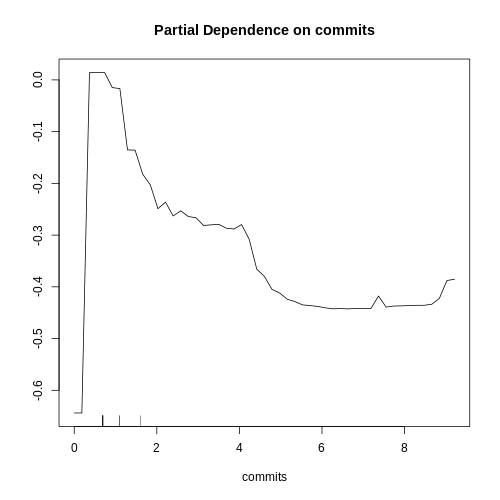} \\
\includegraphics[width = 0.3\linewidth, height = 0.08\textheight]{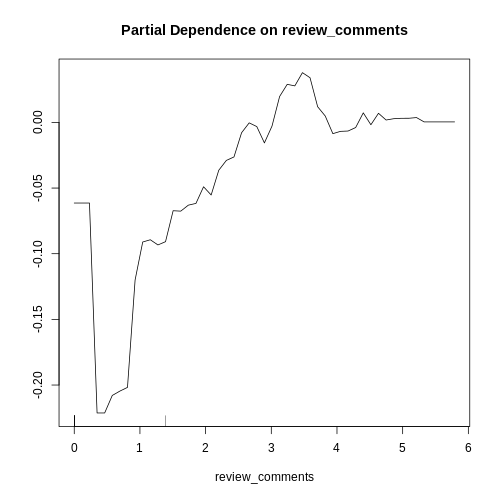} &
\includegraphics[width = 0.3\linewidth, height = 0.08\textheight]{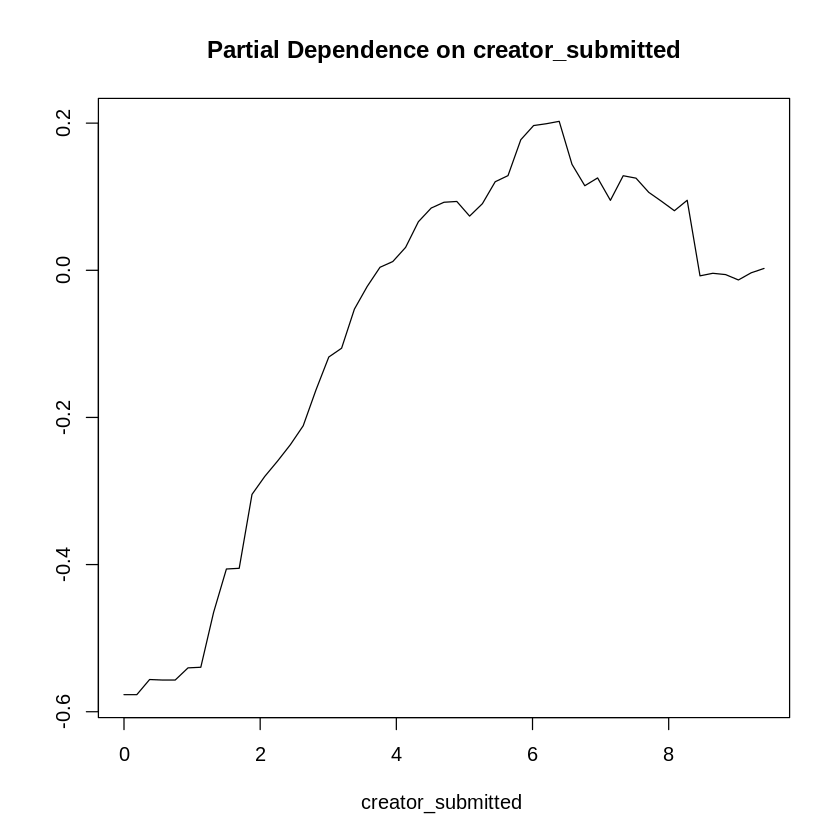} \\ 
\includegraphics[width = 0.3\linewidth, height = 0.08\textheight]{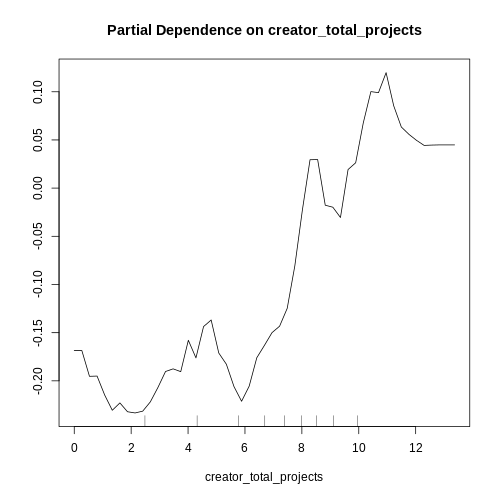} &
\includegraphics[width = 0.3\linewidth, height = 0.08\textheight]{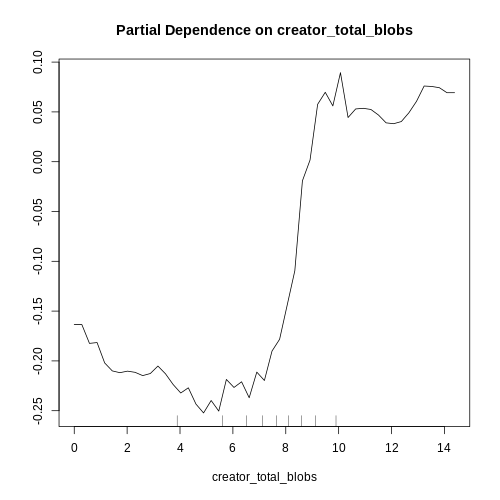} \\
\includegraphics[width = 0.3\linewidth, height = 0.08\textheight]{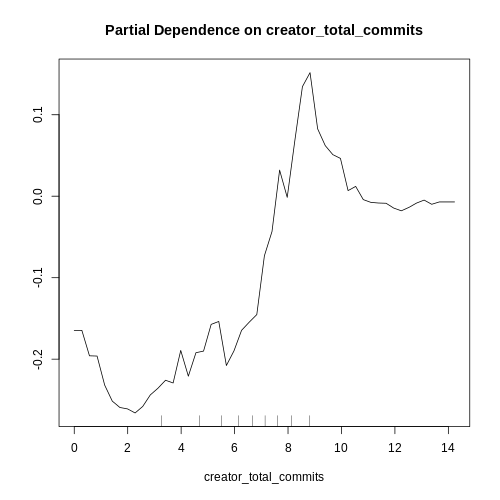}  & 
\includegraphics[width = 0.3\linewidth, height = 0.08\textheight]{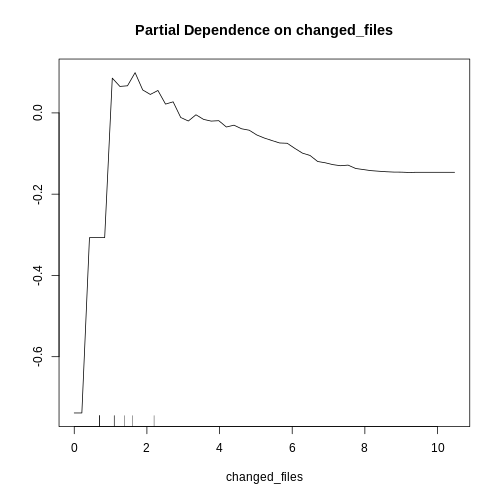} 
\end{tabular}%
}
\end{table}

\vspace{-10pt}
\subsection{Assessing the performance of model using the predictors listed in~\cite{prioritizer} (G1)}

As mentioned in Section~\ref{s:method}, we used a tuned Random Forest model and assessing its predictive performance using 13 out of the 14 predictors. The model performed pretty well using these predictors, with the value of AUC under the ROC curve being 0.89.  
By examining the confusion matrix of the predicted result we found that the model has an accuracy of 0.82, the value of Cohen's kappa coefficient was 0.62 , and the values of sensitivity (recall) and specificity (precision) were 0.69 and 0.90 respectively.

The model displayed a lower accuracy than what was reported in~\cite{prioritizer} using a Random Forest model, which is not surprising, since we are looking at a different research question, and also used one less predictor, for the reasons mentioned in Section~\ref{s:method}. 

The variable importance plot for the model is shown in Figure~\ref{fig:rf1}. We used the same variable names as in~\cite{prioritizer} for ease of interpretation. 

\hypobox{We get a pretty good result using the predictors listed in~\cite{prioritizer}, with AUC-ROC of 0.89}

\vspace{-10pt}
\subsection{Assessing the performance of the model using our set of predictors (G2)}

When we used the 14 predictors as described in Section~\ref{s:data}, we had a significant improvement in result. The tuned Random forest model gave an AUC-ROC value of 0.95. 
By examining the confusion matrix of the predicted result we found that the model has an accuracy of 0.88, the value of Cohen's kappa coefficient was 0.75 , and the values of sensitivity (recall) and specificity (precision) were 0.78 and 0.95 respectively. The variable importance plot for the model is shown in Figure~\ref{fig:rf2}. This is the result that answers our overarching research question when no constraints are put on the problem.

\hypobox{The performance of the model improved significantly with our set of 14 predictors, with AUC-ROC of 0.95}

\vspace{-10pt}
\subsection{Understand the effect of each predictor on the probability of the PR getting merged (G3)}

To illustrate the effects the different predictors have on our model, we looked at their partial dependence plots. A partial dependence plot gives a graphical depiction of the marginal effect of a variable on the class probability for the task of classification, and is implemented by the \textit{partialPlot} function in the \textit{randomForest} R package.
In the X axes of a plots we have the values of the variables (log transformed values in our case, since the variables were log transformed), and the Y axes of the plots show the relative logit contribution of the variable on the class probability~\cite{friedman2001greedy} (probability that a PR was merged, in our case) from the perspective of the model, i.e. negative values (in the Y-axis) mean that the positive class is less likely (i.e. it is less likely that a PR would be accepted, in our case)  for that value of the independent variable (X-axis) according to the model and vice versa. These plots can shed light into the dynamics of PR creation and acceptance, and would be helpful for both the PR creators and the integrators for understanding how to improve the quality of PRs being submitted and accepted. 

The partial dependence plots for all the 14 predictors variables used the Random Forest model created for addressing G2 are shown in Table~\ref{t:pplot} (we apologize for the small font size). The values of all variables shown in the X-axis, except the two fractions and \textit{age}, are in log scale. The values of the two fractions, \textit{creator\_accepted} and \textit{repo\_accepted} are shown in absolute value, between 0 and 1, and the value of age is shown in days, in linear scale.

\hypobox{The marginal effects of the predictor variables on the probability of PR acceptance are shown in Table~\ref{t:pplot}}

\vspace{-10pt}
\subsection{Assessing the performance of the model without using the time dependent predictors (G4)}

Out of the 14 variables we have, the number of discussion comments, number of code review comments, and the age of the PR are variables that vary during the PR's lifetime. Other variables are either historical measures, based on the PR creator's or the repositories activities prior to submission of the PR, or they are static variables that do not change with time. Therefore, to address the goal G4,  we removed the number of discussion comments, the number of code review comments, and the age of the PR from our list of predictors. 

Unsurprisingly, the model with the remaining 11 predictors gave a relatively worse performance, with an AUC-ROC value of 0.89. The confusion matrix revealed that the model has an accuracy of 0.79, the value of Cohen's kappa coefficient was 0.56, and the values of sensitivity (recall) and specificity (precision) were 0.67 and 0.87 respectively. The variable importance plot for the model is shown in Figure~\ref{fig:rf4}.

\hypobox{Using only the predictors available at the time of PR submission, we still get a very good result, with AUC-ROC of 0.89}

\vspace{-10pt}
\subsection{An illustrative example of an NPM package on which this model was applied to test its viability (G5)}

As discussed in Sections~\ref{s:method} and~\ref{s:data}, we used the historical data of the ``bootstrap'' NPM package to predict if the PRs created in future were accepted in an attempt to test the practical viability of our approach.  In the first attempt, we used our full set of 14 predictors for modeling. Although the training and the test sets were somewhat unbalanced in terms of fraction of PRs accepted (see Section~\ref{s:data}), the model performed very well, giving an AUC-ROC value of  0.94, and the confusion matrix showed that it has an accuracy of 0.87, the value of Cohen's kappa coefficient was 0.74, and the values of sensitivity (recall) and specificity (precision) were 0.82 and 0.92 respectively. The variable importance plot for this model is shown in Figure~\ref{fig:rf51}.

We also wanted to check how our the model performs under a condition like G4, so in the second attempt, we tested with the 11 predictors as we did when addressing G4. The model performed fairly, with a modest AUC-ROC value of 0.77. By examining the confusion matrix we found that the model has an accuracy of 0.72, the value of Cohen's kappa coefficient was 0.41 , and the values of sensitivity (recall) and specificity (precision) were 0.54 and 0.85 respectively. The variable importance plot for this model is shown in Figure~\ref{fig:rf52}.

\hypobox{For NPM package ``bootstrap'', training the model with historical data and trying to predict PR acceptance in future gives very good result with AUC-ROC of 0.94 with all predictors, and gives a fair performance with AUC-ROC of 0.77 with only the predictors available at PR submission time}


\section{Discussion}\label{s:disc}

\begin{figure}[!t]
\centering
\vspace{-10pt}
\includegraphics[width=0.6\linewidth]{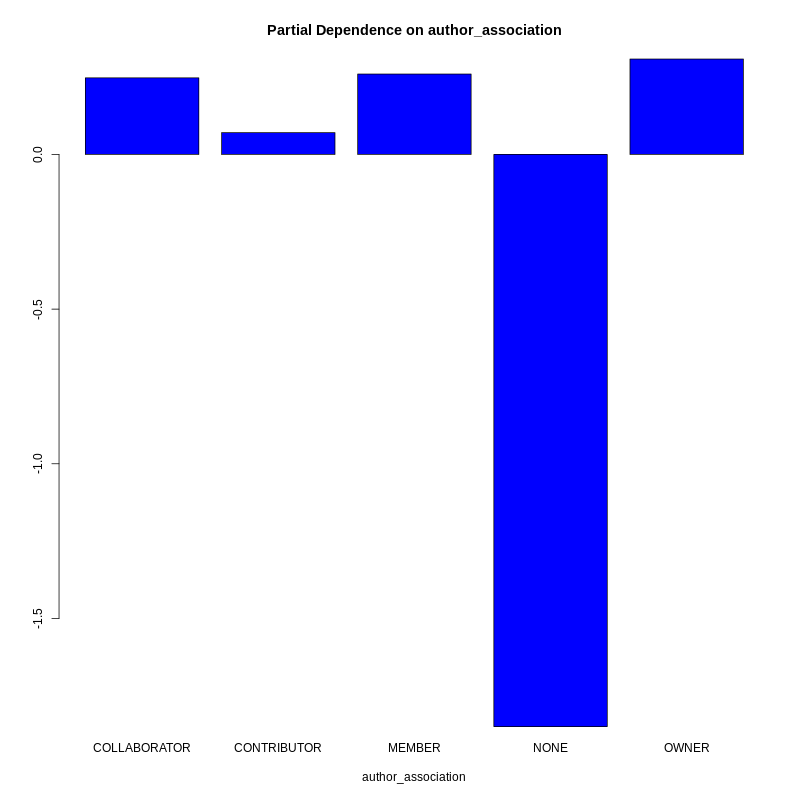}%
\vspace{-10pt}
\caption{Partial Dependence on author association, from the first Random Forest model we made with all 50 predictors}
\label{fig:pdaa}
\vspace{-20pt}
\end{figure}


First, why the 14 predictors we proposed have resulted in much more
accurate predictions than the 13\footnote{we used 13 out of the 14}
predictors used in~\cite{prioritizer}. Although eight variables were
shared between the two sets of predictors, we achieved a much better
performance by adding six other variables: two variables characterizing the
historic PR submission and acceptance rates of the repository in
question, one variable specifying the total number of PRs submitted
by the creator (prior to the predicted PR), and three variables
characterizing the creators' overall activity across all open source
projects that use Git.  In particular, the number of past PRs and the fraction of
those that were accepted (both for the repository to which the PR was submitted,
and for the PR creator across all NPM packages under consideration)
were very important, as we can see by looking at the variable
importance plots. The other variables used in~\cite{prioritizer}
were found not to improve the prediction performance.

We found the age of the PR to consistently show up as the most
significant predictor, and that the decision to merge most PRs is made very
rapidly. Among the PRs that were accepted, $\sim75\%$ are accepted
within 3 days (see the partial dependence plot on \textit{age} in
Table~\ref{t:pplot}), while more than half of the PRs that are not
accepted are closed within 10 days. This is visible from the partial
dependence plot for \textit{age} as well, which shows that the
chance of a PR getting accept rises rapidly during the first day,
and then drops very fast over time. 
This suggests that most NPM package integrators are very responsive, and efficient in handling PRs.

This coupled with the fact that around 60\% of the submitted PRs are
accepted indicate that the NPM ecosystem might have a relatively
lax requirement for accepting a PR. However, when we looked
at the PR creators' association with the projects to which the PR
was submitted, we found that being a member of the organization that
owns the repository, or being the owner of the repository, or having
been invited to collaborate on the repository significantly
increases the probability of a creator's PR(s) being accepted, while
having no connection to the repository decreases the probability
significantly (see Figure~\ref{fig:pdaa}). This result,
unfortunately, could be misleading, since the membership information was
obtained not at the time the PR was accepted or created (like all
other predictors we use) but much later, at the time of data
collection, and we have no way to verify that the 
author's affiliation wasn't updated retroactively.
Our suspicion was reinforced by the finding that for the ``bootstrap''
project, no PR creator's association was updated over the lifetime
of the project, i.e. a PR creator who's a \textit{contributor} was 
found to always have been a contributor to the project, since the 
very first PR they created. A similar situation was later observed for
all the NPM projects, that an author's association with a particular
project never seem to change in the data that we collected.
This is the reason why this variable, in spite of
appearing to be an important predictor (a variation of this variable,
viz. \textit{Core Member} was used in~\cite{prioritizer} as well),
wasn't included in our model (see Section~\ref{s:data}). In a
nutshell, it might include information that was not available when the
decision to merge the PR was made. 

We also found that the historical performance record of the PR
creator, in terms of what number of the PRs created by that
individual and their acceptance rate before submitting this PR, and
the number of PRs submitted to the repository and the acceptance
rate of those PRs are the other most significant predictors. Looking
at the partial dependence plots for these variables in
Table~\ref{t:pplot}, we see that:
\vspace{-5pt}
\begin{itemize}
    \item  As the number of PRs submitted by the PR creator, prior
      to submitting the predicted PR, increases, the probability of
      acceptance of the PR submitted by that creator increases up
      to a certain point, and then it flattens out, indicating the
      presence of a saturation point in the author/project relationship.
    \item The probability that a PR submitted by a creator will be
      accepted has a strong positive correlation with the fraction
      of PRs (created by the same creator) accepted prior to
      submitting that PR, indicating a strong dependence on the
      creator's proficiency.  
    \item We also see that repositories that accept a larger
      fraction of the PRs submitted are likely to accept more PRs,
      which could indicate a more lenient policy of PR acceptance
      and/or the integrators' willingness to accept more PRs. 
    \item We see a somewhat surprising result on how the chances of
      a PR being accepted varies with the number of PRs submitted
      against the repository. We see that the repositories that
      receive a moderate number PRs are less likely to accept
      PRs. The reason why the repositories with few PRs are more
      likely to accept could be because they are relatively new
      and/or the integrators aren't too busy. The reason why
      repositories with more PRs are more likely to accept could
      boil down to the type of the project, the leniency in policy,
      higher efficiency of integrators,
      and/or the knowledge among the PR creators that it is easier to
      get a PR accepted in those repositories. 
\end{itemize}
\vspace{-5pt}

Another observation can be made by looking at the partial
dependence plots of different properties of the PRs, like
\textit{additions, deletions, commits, changed\_files} in
Table~\ref{t:pplot}, that smaller patches are more likely to be
accepted, which reinforces the findings
of~\cite{weissgerber2008small}. However, we see that PRs with very
few changes are unlikely to be accepted, possibly because the
contribution isn't sufficient. These plots indicate the presence
of a sort of \textit{sweet spot} or \textit{Goldilocks Zone} in
terms of the size of the patch, with around 1-4 commits, 30-250
lines deleted, 5-100 lines added, and 5-15 files changed. Our findings
are similar to what was reported in~\cite{soares2015acceptance}.

\hypobox{
Pull Requests with around 1--4 commits, 30-250 lines deleted, 5-100 lines added, and 5-15 files changed have a much higher chance of getting accepted.
}

In terms of \textit{comments} and \textit{review\_comments}, we see
that having none or very few means a PR will likely be accepted, but
a few more indicate that it will most likely be rejected, and having
even more \textit{comments} and especially \textit{review\_comments}
again increases the chance of acceptance. This could indicate a
pattern in the discussion boards, where the good, and likely simpler
PRs are accepted with few or no discussion, while the more
controversial PRs invite a few more comments, likely explaining why
it wasn't accepted, and the more complex PRs tend to have a more
elaborate discussion but still have a fair chance of acceptance.

If we look at how the PR creators' activity across all OSS projects
using Git affect their chance of getting a PR accepted, we can
clearly see the presence of a threshold, after crossing which the
creators' PRs have a much higher chance of getting accepted. Looking
at the plots in Table~\ref{t:pplot} for these three variables, we
approximate the threshold to be around 700 projects, 1000 commits,
and 3000 blobs. Developers who have worked in around 700 or more
projects, made around 1000 or commits, and authored around 3000 or
more blobs seem to have achieved a level of proficiency or
reputation that significantly improves the chances that the PRs they
submit get merged.

A number of variables didn't make it to the list of the useful
predictors. We see that most of the characteristics of the head and
base repositories do not help in increasing accuracy, including the number of stars,
forks etc., which are generally accepted as the measures of
popularity or reputation (see e.g.~\cite{borges2016understanding}), and the number
of issues, which should be an important predictor intuitively, since
more issues might invite more PRs that try to fix those
issues. Moreover, if a PR was submitted with the fix for an issue
($\sim43\%$ of the PRs have an issue fix) also seem to be not very
significant.

\vspace{-10pt}
\section{Limitations}\label{s:limit}

Regarding external validity, we looked at 4218 most popular NPM packages, which,
while a large number by itself, is less than 0.5\% of the total
packages in the NPM ecosystem. These packages, however, represent
the tiny part of the NPM ecosystem that is widely used and where the
vast majority of code contribution actually happens. 

We extracted 50 variables from the data we obtained, which should
cover most of the latent factors affecting the chances of a PR being
accepted, however, it may not be an exhaustive set
of variables. There may still be additional variables that either
improve the predictor performance or help explain  why certain PRs
are more likely to be accepted. 


Finally, the result of this paper might not be applicable as-is to other software ecosystems, since every ecosystem has their norms and characteristics which is impossible to account for when looking into only one ecosystem. Future
studies are needed to determine the generality of our findings.

\vspace{-10pt}
\section{Conclusion}\label{s:conclusion}

First, we have answered the overarching research question by
determining that, at least for popular NPM packages (the ones with 
more than 10,000 monthly downloads), we can predict
extremely accurately if a PR will be merged within a month of being
created (AUC-ROC of 0.95), and illustrated the response curve to the
values of the key predictors that can be used to understand the
rather specific characteristics of the PRs that are most likely to
get accepted. We were able to achieve such high precision by
looking at the historical PR submission and acceptance records 
of the repositories to which a PR was submitted, and by
incorporating information about developer activity across unrelated software projects,
specifically using it as a proxy for the experience or reputation of
the authors. This was made possible by exploiting data collected
from the entire OSS ecosystem~\cite{woc19} and it significantly
improved upon the results obtained by using the predictors listed
in~\cite{prioritizer}.

We have also explored the practical aspects of such prediction:
would it work if we only use the properties of the PR available at
the time of the creation (in contrast to using the properties of the
PR just before it is merged) and found the prediction to be still
highly accurate. Finally we evaluate how accurate the approach
would be if it was applied in specific large NPM package.

Our findings have theoretical and practical implications. The
accuracy of models of PR acceptance increase the likelihood of
successful practical applications that range from tools that
support PR integrators to tools that help authors of the PRs to
tailor their contributions to the form resembling that of the PRs
that most likely to be accepted by a specific project. We plan to
pursue the goal of evaluating such tools in OSS projects. As the NPM
ecosystem and other OSS ecosystems depend on contributors to
maintain growth and code quality, we hope that the results of our
work would help these ecosystems to sustain evolution and high
quality of the code.

\balance
\bibliographystyle{ACM-Reference-Format}
\bibliography{sigproc} 


\begin{thebibliography}{00}


\ifx \showCODEN    \undefined \def \showCODEN     #1{\unskip}     \fi
\ifx \showDOI      \undefined \def \showDOI       #1{#1}\fi
\ifx \showISBNx    \undefined \def \showISBNx     #1{\unskip}     \fi
\ifx \showISBNxiii \undefined \def \showISBNxiii  #1{\unskip}     \fi
\ifx \showISSN     \undefined \def \showISSN      #1{\unskip}     \fi
\ifx \showLCCN     \undefined \def \showLCCN      #1{\unskip}     \fi
\ifx \shownote     \undefined \def \shownote      #1{#1}          \fi
\ifx \showarticletitle \undefined \def \showarticletitle #1{#1}   \fi
\ifx \showURL      \undefined \def \showURL       {\relax}        \fi
\providecommand\bibfield[2]{#2}
\providecommand\bibinfo[2]{#2}
\providecommand\natexlab[1]{#1}
\providecommand\showeprint[2][]{arXiv:#2}

\bibitem[\protect\citeauthoryear{Baysal, Kononenko, Holmes, and Godfrey}{Baysal
  et~al\mbox{.}}{2012}]%
        {baysal2012secret}
\bibfield{author}{\bibinfo{person}{Olga Baysal}, \bibinfo{person}{Oleksii
  Kononenko}, \bibinfo{person}{Reid Holmes}, {and} \bibinfo{person}{Michael~W
  Godfrey}.} \bibinfo{year}{2012}\natexlab{}.
\newblock \showarticletitle{The secret life of patches: A firefox case study}.
  In \bibinfo{booktitle}{{\em 2012 19th Working Conference on Reverse
  Engineering}}. IEEE, \bibinfo{pages}{447--455}.
\newblock


\bibitem[\protect\citeauthoryear{Bleikamp}{Bleikamp}{2012}]%
        {GHPR}
\bibfield{author}{\bibinfo{person}{Ben Bleikamp}.}
  \bibinfo{year}{2012}\natexlab{}.
\newblock \bibinfo{title}{How we use Pull Requests to build GitHub}.
\newblock   (\bibinfo{year}{2012}).
\newblock
\showURL{%
\url{https://github.blog/2012-05-02-how-we-use-pull-requests-to-build-github/}}


\bibitem[\protect\citeauthoryear{Borges, Hora, and Valente}{Borges
  et~al\mbox{.}}{2016}]%
        {borges2016understanding}
\bibfield{author}{\bibinfo{person}{Hudson Borges}, \bibinfo{person}{Andre
  Hora}, {and} \bibinfo{person}{Marco~Tulio Valente}.}
  \bibinfo{year}{2016}\natexlab{}.
\newblock \showarticletitle{Understanding the factors that impact the
  popularity of GitHub repositories}. In \bibinfo{booktitle}{{\em Software
  Maintenance and Evolution (ICSME), 2016 IEEE International Conference on}}.
  IEEE, \bibinfo{pages}{334--344}.
\newblock


\bibitem[\protect\citeauthoryear{Dabbish, Stuart, Tsay, and Herbsleb}{Dabbish
  et~al\mbox{.}}{2012}]%
        {dabbish2012social}
\bibfield{author}{\bibinfo{person}{Laura Dabbish}, \bibinfo{person}{Colleen
  Stuart}, \bibinfo{person}{Jason Tsay}, {and} \bibinfo{person}{Jim Herbsleb}.}
  \bibinfo{year}{2012}\natexlab{}.
\newblock \showarticletitle{Social coding in GitHub: transparency and
  collaboration in an open software repository}. In \bibinfo{booktitle}{{\em
  Proceedings of the ACM 2012 conference on computer supported cooperative
  work}}. ACM, \bibinfo{pages}{1277--1286}.
\newblock


\bibitem[\protect\citeauthoryear{Decan, Mens, and Constantinou}{Decan
  et~al\mbox{.}}{2018}]%
        {decan2018impact}
\bibfield{author}{\bibinfo{person}{Alexandre Decan}, \bibinfo{person}{Tom
  Mens}, {and} \bibinfo{person}{Eleni Constantinou}.}
  \bibinfo{year}{2018}\natexlab{}.
\newblock \showarticletitle{On the impact of security vulnerabilities in the
  npm package dependency network}. In \bibinfo{booktitle}{{\em 2018 IEEE/ACM
  15th International Conference on Mining Software Repositories (MSR)}}. IEEE,
  \bibinfo{pages}{181--191}.
\newblock


\bibitem[\protect\citeauthoryear{Dey, Ma, and Mockus}{Dey
  et~al\mbox{.}}{2019}]%
        {dey2019patterns}
\bibfield{author}{\bibinfo{person}{Tapajit Dey}, \bibinfo{person}{Yuxing Ma},
  {and} \bibinfo{person}{Audris Mockus}.} \bibinfo{year}{2019}\natexlab{}.
\newblock \showarticletitle{Patterns of Effort Contribution and Demand and User
  Classification based on Participation Patterns in NPM Ecosystem}.
\newblock \bibinfo{journal}{{\em arXiv preprint arXiv:1907.06538\/}}
  (\bibinfo{year}{2019}).
\newblock


\bibitem[\protect\citeauthoryear{Dey and Mockus}{Dey and Mockus}{2018}]%
        {dey2018software}
\bibfield{author}{\bibinfo{person}{Tapajit Dey} {and} \bibinfo{person}{Audris
  Mockus}.} \bibinfo{year}{2018}\natexlab{}.
\newblock \showarticletitle{Are software dependency supply chain metrics useful
  in predicting change of popularity of npm packages?}. In
  \bibinfo{booktitle}{{\em Proceedings of the 14th International Conference on
  Predictive Models and Data Analytics in Software Engineering}}. ACM,
  \bibinfo{pages}{66--69}.
\newblock


\bibitem[\protect\citeauthoryear{Friedman}{Friedman}{2001}]%
        {friedman2001greedy}
\bibfield{author}{\bibinfo{person}{Jerome~H Friedman}.}
  \bibinfo{year}{2001}\natexlab{}.
\newblock \showarticletitle{Greedy function approximation: a gradient boosting
  machine}.
\newblock \bibinfo{journal}{{\em Annals of statistics\/}}
  (\bibinfo{year}{2001}), \bibinfo{pages}{1189--1232}.
\newblock


\bibitem[\protect\citeauthoryear{Gousios}{Gousios}{2013}]%
        {Ghtorrent}
\bibfield{author}{\bibinfo{person}{Georgios Gousios}.}
  \bibinfo{year}{2013}\natexlab{}.
\newblock \showarticletitle{The GHTorrent dataset and tool suite}. In
  \bibinfo{booktitle}{{\em Proceedings of the 10th Working Conference on Mining
  Software Repositories}} {\em (\bibinfo{series}{MSR '13})}.
  \bibinfo{publisher}{IEEE Press}, \bibinfo{address}{Piscataway, NJ, USA},
  \bibinfo{pages}{233--236}.
\newblock
\showISBNx{978-1-4673-2936-1}
\showURL{%
\url{http://dl.acm.org/citation.cfm?id=2487085.2487132}}


\bibitem[\protect\citeauthoryear{Gousios, Pinzger, and Deursen}{Gousios
  et~al\mbox{.}}{2014}]%
        {gousios2014exploratory}
\bibfield{author}{\bibinfo{person}{Georgios Gousios}, \bibinfo{person}{Martin
  Pinzger}, {and} \bibinfo{person}{Arie~van Deursen}.}
  \bibinfo{year}{2014}\natexlab{}.
\newblock \showarticletitle{An exploratory study of the pull-based software
  development model}. In \bibinfo{booktitle}{{\em Proceedings of the 36th
  International Conference on Software Engineering}}. ACM,
  \bibinfo{pages}{345--355}.
\newblock


\bibitem[\protect\citeauthoryear{Gousios, Storey, and Bacchelli}{Gousios
  et~al\mbox{.}}{2016}]%
        {gousios2016work}
\bibfield{author}{\bibinfo{person}{Georgios Gousios},
  \bibinfo{person}{Margaret-Anne Storey}, {and} \bibinfo{person}{Alberto
  Bacchelli}.} \bibinfo{year}{2016}\natexlab{}.
\newblock \showarticletitle{Work practices and challenges in pull-based
  development: the contributor's perspective}. In \bibinfo{booktitle}{{\em 2016
  IEEE/ACM 38th International Conference on Software Engineering (ICSE)}}.
  IEEE, \bibinfo{pages}{285--296}.
\newblock


\bibitem[\protect\citeauthoryear{Gousios, Zaidman, Storey, and
  Van~Deursen}{Gousios et~al\mbox{.}}{2015}]%
        {gousios2015work}
\bibfield{author}{\bibinfo{person}{Georgios Gousios}, \bibinfo{person}{Andy
  Zaidman}, \bibinfo{person}{Margaret-Anne Storey}, {and} \bibinfo{person}{Arie
  Van~Deursen}.} \bibinfo{year}{2015}\natexlab{}.
\newblock \showarticletitle{Work practices and challenges in pull-based
  development: the integrator's perspective}. In \bibinfo{booktitle}{{\em
  Proceedings of the 37th International Conference on Software
  Engineering-Volume 1}}. IEEE Press, \bibinfo{pages}{358--368}.
\newblock


\bibitem[\protect\citeauthoryear{Jiang, Yang, He, Blanc, and Zhang}{Jiang
  et~al\mbox{.}}{2017}]%
        {jiang2017should}
\bibfield{author}{\bibinfo{person}{Jing Jiang}, \bibinfo{person}{Yun Yang},
  \bibinfo{person}{Jiahuan He}, \bibinfo{person}{Xavier Blanc}, {and}
  \bibinfo{person}{Li Zhang}.} \bibinfo{year}{2017}\natexlab{}.
\newblock \showarticletitle{Who should comment on this pull request? analyzing
  attributes for more accurate commenter recommendation in pull-based
  development}.
\newblock \bibinfo{journal}{{\em Information and Software Technology\/}}
  \bibinfo{volume}{84} (\bibinfo{year}{2017}), \bibinfo{pages}{48--62}.
\newblock


\bibitem[\protect\citeauthoryear{Jiang, Adams, and German}{Jiang
  et~al\mbox{.}}{2013}]%
        {jiang2013will}
\bibfield{author}{\bibinfo{person}{Yujuan Jiang}, \bibinfo{person}{Bram Adams},
  {and} \bibinfo{person}{Daniel~M German}.} \bibinfo{year}{2013}\natexlab{}.
\newblock \showarticletitle{Will my patch make it? and how fast?: Case study on
  the linux kernel}. In \bibinfo{booktitle}{{\em Proceedings of the 10th
  Working Conference on Mining Software Repositories}}. IEEE Press,
  \bibinfo{pages}{101--110}.
\newblock


\bibitem[\protect\citeauthoryear{Ma, Bogart, Amreen, Zaretzki, and Mockus}{Ma
  et~al\mbox{.}}{2019}]%
        {woc19}
\bibfield{author}{\bibinfo{person}{Yuxing Ma}, \bibinfo{person}{Chris Bogart},
  \bibinfo{person}{Sadika Amreen}, \bibinfo{person}{Russell Zaretzki}, {and}
  \bibinfo{person}{Audris Mockus}.} \bibinfo{year}{2019}\natexlab{}.
\newblock \showarticletitle{World of Code: An Infrastructure for Mining the
  Universe of Open Source VCS Data}. In \bibinfo{booktitle}{{\em IEEE Working
  Conference on Mining Software Repositories}}.
\newblock
\showURL{%
\url{papers/WoC.pdf}}


\bibitem[\protect\citeauthoryear{Mockus, Fielding, and Herbsleb}{Mockus
  et~al\mbox{.}}{2002}]%
        {mockus2002two}
\bibfield{author}{\bibinfo{person}{Audris Mockus}, \bibinfo{person}{Roy~T
  Fielding}, {and} \bibinfo{person}{James~D Herbsleb}.}
  \bibinfo{year}{2002}\natexlab{}.
\newblock \showarticletitle{Two case studies of open source software
  development: Apache and Mozilla}.
\newblock \bibinfo{journal}{{\em ACM Transactions on Software Engineering and
  Methodology (TOSEM)\/}} \bibinfo{volume}{11}, \bibinfo{number}{3}
  (\bibinfo{year}{2002}), \bibinfo{pages}{309--346}.
\newblock


\bibitem[\protect\citeauthoryear{Rahman and Roy}{Rahman and Roy}{2014}]%
        {rahman2014insight}
\bibfield{author}{\bibinfo{person}{Mohammad~Masudur Rahman} {and}
  \bibinfo{person}{Chanchal~K Roy}.} \bibinfo{year}{2014}\natexlab{}.
\newblock \showarticletitle{An insight into the pull requests of github}. In
  \bibinfo{booktitle}{{\em Proceedings of the 11th Working Conference on Mining
  Software Repositories}}. ACM, \bibinfo{pages}{364--367}.
\newblock


\bibitem[\protect\citeauthoryear{Rigby, German, Cowen, and Storey}{Rigby
  et~al\mbox{.}}{2014}]%
        {rigby2014peer}
\bibfield{author}{\bibinfo{person}{Peter~C Rigby}, \bibinfo{person}{Daniel~M
  German}, \bibinfo{person}{Laura Cowen}, {and} \bibinfo{person}{Margaret-Anne
  Storey}.} \bibinfo{year}{2014}\natexlab{}.
\newblock \showarticletitle{Peer review on open-source software projects:
  Parameters, statistical models, and theory}.
\newblock \bibinfo{journal}{{\em ACM Transactions on Software Engineering and
  Methodology (TOSEM)\/}} \bibinfo{volume}{23}, \bibinfo{number}{4}
  (\bibinfo{year}{2014}), \bibinfo{pages}{35}.
\newblock


\bibitem[\protect\citeauthoryear{Soares, de~Lima~J{\'u}nior, Murta, and
  Plastino}{Soares et~al\mbox{.}}{2015}]%
        {soares2015acceptance}
\bibfield{author}{\bibinfo{person}{Daric{\'e}lio~Moreira Soares},
  \bibinfo{person}{Manoel~Limeira de Lima~J{\'u}nior},
  \bibinfo{person}{Leonardo Murta}, {and} \bibinfo{person}{Alexandre
  Plastino}.} \bibinfo{year}{2015}\natexlab{}.
\newblock \showarticletitle{Acceptance factors of pull requests in open-source
  projects}. In \bibinfo{booktitle}{{\em Proceedings of the 30th Annual ACM
  Symposium on Applied Computing}}. ACM, \bibinfo{pages}{1541--1546}.
\newblock


\bibitem[\protect\citeauthoryear{Tsay, Dabbish, and Herbsleb}{Tsay
  et~al\mbox{.}}{2014}]%
        {tsay2014influence}
\bibfield{author}{\bibinfo{person}{Jason Tsay}, \bibinfo{person}{Laura
  Dabbish}, {and} \bibinfo{person}{James Herbsleb}.}
  \bibinfo{year}{2014}\natexlab{}.
\newblock \showarticletitle{Influence of social and technical factors for
  evaluating contribution in GitHub}. In \bibinfo{booktitle}{{\em Proceedings
  of the 36th international conference on Software engineering}}. ACM,
  \bibinfo{pages}{356--366}.
\newblock


\bibitem[\protect\citeauthoryear{Tu, Zhu, Zheng, and Zhou}{Tu
  et~al\mbox{.}}{2018}]%
        {tu2018careful}
\bibfield{author}{\bibinfo{person}{Feifei Tu}, \bibinfo{person}{Jiaxin Zhu},
  \bibinfo{person}{Qimu Zheng}, {and} \bibinfo{person}{Minghui Zhou}.}
  \bibinfo{year}{2018}\natexlab{}.
\newblock \showarticletitle{Be careful of when: an empirical study on
  time-related misuse of issue tracking data}. In \bibinfo{booktitle}{{\em
  Proceedings of the 2018 26th ACM Joint Meeting on European Software
  Engineering Conference and Symposium on the Foundations of Software
  Engineering}}. ACM, \bibinfo{pages}{307--318}.
\newblock


\bibitem[\protect\citeauthoryear{v.~d. {Veen}, {Gousios}, and {Zaidman}}{v.~d.
  {Veen} et~al\mbox{.}}{2015}]%
        {prioritizer}
\bibfield{author}{\bibinfo{person}{E. v.~d. {Veen}}, \bibinfo{person}{G.
  {Gousios}}, {and} \bibinfo{person}{A. {Zaidman}}.}
  \bibinfo{year}{2015}\natexlab{}.
\newblock \showarticletitle{Automatically Prioritizing Pull Requests}. In
  \bibinfo{booktitle}{{\em 2015 IEEE/ACM 12th Working Conference on Mining
  Software Repositories}}. \bibinfo{pages}{357--361}.
\newblock
\showISSN{2160-1852}
\showDOI{%
\url{https://doi.org/10.1109/MSR.2015.40}}


\bibitem[\protect\citeauthoryear{Voss}{Voss}{2016}]%
        {npmuser}
\bibfield{author}{\bibinfo{person}{Laurie Voss}.}
  \bibinfo{year}{2016}\natexlab{}.
\newblock \bibinfo{title}{how many npm users are there?}
\newblock   (\bibinfo{year}{2016}).
\newblock
\showURL{%
\url{https://blog.npmjs.org/post/143451680695/how-many-npm-users-are-there}}


\bibitem[\protect\citeauthoryear{Wei{\ss}gerber, Neu, and Diehl}{Wei{\ss}gerber
  et~al\mbox{.}}{2008}]%
        {weissgerber2008small}
\bibfield{author}{\bibinfo{person}{Peter Wei{\ss}gerber},
  \bibinfo{person}{Daniel Neu}, {and} \bibinfo{person}{Stephan Diehl}.}
  \bibinfo{year}{2008}\natexlab{}.
\newblock \showarticletitle{Small patches get in!}. In \bibinfo{booktitle}{{\em
  Proceedings of the 2008 international working conference on Mining software
  repositories}}. ACM, \bibinfo{pages}{67--76}.
\newblock


\bibitem[\protect\citeauthoryear{Wittern, Suter, and Rajagopalan}{Wittern
  et~al\mbox{.}}{2016}]%
        {wittern2016look}
\bibfield{author}{\bibinfo{person}{Erik Wittern}, \bibinfo{person}{Philippe
  Suter}, {and} \bibinfo{person}{Shriram Rajagopalan}.}
  \bibinfo{year}{2016}\natexlab{}.
\newblock \showarticletitle{A look at the dynamics of the JavaScript package
  ecosystem}. In \bibinfo{booktitle}{{\em Mining Software Repositories (MSR),
  2016 IEEE/ACM 13th Working Conference on}}. IEEE, \bibinfo{pages}{351--361}.
\newblock


\bibitem[\protect\citeauthoryear{Xie, Zhou, and Mockus}{Xie
  et~al\mbox{.}}{2013}]%
        {xie2013impact}
\bibfield{author}{\bibinfo{person}{Jialiang Xie}, \bibinfo{person}{Minghui
  Zhou}, {and} \bibinfo{person}{Audris Mockus}.}
  \bibinfo{year}{2013}\natexlab{}.
\newblock \showarticletitle{Impact of triage: a study of mozilla and gnome}. In
  \bibinfo{booktitle}{{\em 2013 ACM/IEEE International Symposium on Empirical
  Software Engineering and Measurement}}. IEEE, \bibinfo{pages}{247--250}.
\newblock


\bibitem[\protect\citeauthoryear{Yu, Wang, Filkov, Devanbu, and Vasilescu}{Yu
  et~al\mbox{.}}{2015}]%
        {yu2015wait}
\bibfield{author}{\bibinfo{person}{Yue Yu}, \bibinfo{person}{Huaimin Wang},
  \bibinfo{person}{Vladimir Filkov}, \bibinfo{person}{Premkumar Devanbu}, {and}
  \bibinfo{person}{Bogdan Vasilescu}.} \bibinfo{year}{2015}\natexlab{}.
\newblock \showarticletitle{Wait for it: determinants of pull request
  evaluation latency on GitHub}. In \bibinfo{booktitle}{{\em 2015 IEEE/ACM 12th
  Working Conference on Mining Software Repositories}}. IEEE,
  \bibinfo{pages}{367--371}.
\newblock


\bibitem[\protect\citeauthoryear{Yu, Wang, Yin, and Ling}{Yu
  et~al\mbox{.}}{2014a}]%
        {yu2014reviewer}
\bibfield{author}{\bibinfo{person}{Yue Yu}, \bibinfo{person}{Huaimin Wang},
  \bibinfo{person}{Gang Yin}, {and} \bibinfo{person}{Charles~X Ling}.}
  \bibinfo{year}{2014}\natexlab{a}.
\newblock \showarticletitle{Reviewer recommender of pull-requests in GitHub}.
  In \bibinfo{booktitle}{{\em 2014 IEEE International Conference on Software
  Maintenance and Evolution}}. IEEE, \bibinfo{pages}{609--612}.
\newblock


\bibitem[\protect\citeauthoryear{Yu, Wang, Yin, and Ling}{Yu
  et~al\mbox{.}}{2014b}]%
        {yu2014should}
\bibfield{author}{\bibinfo{person}{Yue Yu}, \bibinfo{person}{Huaimin Wang},
  \bibinfo{person}{Gang Yin}, {and} \bibinfo{person}{Charles~X Ling}.}
  \bibinfo{year}{2014}\natexlab{b}.
\newblock \showarticletitle{Who should review this pull-request: Reviewer
  recommendation to expedite crowd collaboration}. In \bibinfo{booktitle}{{\em
  2014 21st Asia-Pacific Software Engineering Conference}},
  Vol.~\bibinfo{volume}{1}. IEEE, \bibinfo{pages}{335--342}.
\newblock


\bibitem[\protect\citeauthoryear{Yu, Wang, Yin, and Wang}{Yu
  et~al\mbox{.}}{2016}]%
        {yu2016reviewer}
\bibfield{author}{\bibinfo{person}{Yue Yu}, \bibinfo{person}{Huaimin Wang},
  \bibinfo{person}{Gang Yin}, {and} \bibinfo{person}{Tao Wang}.}
  \bibinfo{year}{2016}\natexlab{}.
\newblock \showarticletitle{Reviewer recommendation for pull-requests in
  GitHub: What can we learn from code review and bug assignment?}
\newblock \bibinfo{journal}{{\em Information and Software Technology\/}}
  \bibinfo{volume}{74} (\bibinfo{year}{2016}), \bibinfo{pages}{204--218}.
\newblock


\bibitem[\protect\citeauthoryear{Zapata, Kula, Chinthanet, Ishio, Matsumoto,
  and Ihara}{Zapata et~al\mbox{.}}{2018}]%
        {zapata2018towards}
\bibfield{author}{\bibinfo{person}{Rodrigo~Elizalde Zapata},
  \bibinfo{person}{Raula~Gaikovina Kula}, \bibinfo{person}{Bodin Chinthanet},
  \bibinfo{person}{Takashi Ishio}, \bibinfo{person}{Kenichi Matsumoto}, {and}
  \bibinfo{person}{Akinori Ihara}.} \bibinfo{year}{2018}\natexlab{}.
\newblock \showarticletitle{Towards smoother library migrations: A look at
  vulnerable dependency migrations at function level for npm JavaScript
  packages}. In \bibinfo{booktitle}{{\em 2018 IEEE International Conference on
  Software Maintenance and Evolution (ICSME)}}. IEEE,
  \bibinfo{pages}{559--563}.
\newblock


\bibitem[\protect\citeauthoryear{Zhu, Zhou, and Mockus}{Zhu
  et~al\mbox{.}}{2016}]%
        {zhu2016effectiveness}
\bibfield{author}{\bibinfo{person}{Jiaxin Zhu}, \bibinfo{person}{Minghui Zhou},
  {and} \bibinfo{person}{Audris Mockus}.} \bibinfo{year}{2016}\natexlab{}.
\newblock \showarticletitle{Effectiveness of code contribution: From
  patch-based to pull-request-based tools}. In \bibinfo{booktitle}{{\em
  Proceedings of the 2016 24th ACM SIGSOFT International Symposium on
  Foundations of Software Engineering}}. ACM, \bibinfo{pages}{871--882}.
\newblock


\end{thebibliography}

\end{document}